%% file: article.tex
\documentclass[final,3p,times,twocolumn]{elsarticle}  

\input{macros.tex}

\usepackage{hyperref}

\usepackage{cleveref}

\newcommand{\PaperTitle}{Analytical Uncertainty Propagation for Multi-Period Stochastic Optimal Power Flow}
\newcommand\scalemath[2]{\scalebox{#1}{\mbox{\ensuremath{\displaystyle #2}}}}

\journal{Journal of \LaTeX\ Templates}

\begin{document}

\begin{frontmatter}

\title{\PaperTitle}

\author[add1]{Rebecca Bauer}
\ead{rebecca.bauer@kit.edu}
\author[add1]{Tillmann~M\"uhlpfordt}
\ead{t.muehlpfordt@mailbox.org}
\author[add2]{Nicole Ludwig}
\ead{nicole.ludwig@uni-tuebingen.de}
\author[add1]{Veit Hagenmeyer}
\ead{veit.hagenmeyer@kit.edu}
\address[add1]{Institute for Automation und applied Informatics, Karlsruhe Institute of Technology}
\address[add2]{Cluster of Excellence “Machine Learning: New Perspectives for Science”, University of T\"ubingen, Germany}

\begin{abstract}

The increase in \glspl{res}, like wind or solar power, results in growing uncertainty also in transmission grids. 
This affects grid stability through fluctuating energy supply and an increased probability of overloaded lines. 
One key strategy to cope with this uncertainty is the use of distributed \glspl{ess}. 
In order to securely operate power systems containing renewables and use storage, optimization models are needed that both handle uncertainty and apply \glspl{ess}.

This paper introduces a compact dynamic stochastic chance-constrained optimal power flow~(CC-OPF) model, that minimizes generation costs and includes distributed \glspl{ess}. Assuming Gaussian uncertainty, we use affine policies to obtain a tractable, analytically exact reformulation as a \gls{socp}.
We test the new model on five different IEEE networks with varying sizes of 5, 39, 57, 118 and 300 nodes and include complexity analysis. 
The results show that the model is computationally efficient and robust with respect to constraint violation risk. The distributed energy storage system leads to more stable operation with flattened generation profiles. Storage absorbed RES uncertainty, and reduced generation cost. 

\end{abstract}


\end{frontmatter}



\section{Introduction}

\glsresetall

Volatile renewable energy resources, such as wind, are increasingly included in power systems. Besides many benefits, these \glspl{res} bring more uncertainty into the system as they depend on fluctuating weather dynamics. This challenges the grid's reliability and leads to frequency fluctuations or RES curtailment. To cope with these new challenges, more and more research focuses on the operation of power systems under uncertainty \cite{Capitanescu12,Vrakopoulou17,Roald18,Guo18,Li17,Warrington13}. 
A central strategy to securely operate power systems under uncertainty is the inclusion of distributed \glspl{ess}. E.g. currently many grid boosters are installed in transmission grids \cite{molina_distributed_2012}; in Germany \cite{figgener_development_2021}, Europe \cite{zsiboracs_intermittent_2019} and in the world \cite{gulagi_role_2018, keck_impact_2019}. 
In contrast to conventional power plants (e.g. thermal, gas), \glspl{ess} have the advantage that it costs less, and it can store and discharge power of renewables. During the ongoing lifetime of thermal power plants it can also react much quicker to fluctuations. 


In order to tackle uncertainty in power systems together with storage, we can use DC \gls{opf}. \Gls{opf} is a standard tool for operating power systems in a way that minimizes operational costs, while respecting both the physical limits of the network such as line flow limits, and the power flow equation for system stability. The DC linearization of AC power flow is a standard approximation method \cite{wood_power_2014}. Under stochastic uncertainty the DC OPF can be formulated as a chance-constrained OPF (CC-OPF), which is exact when assuming Gaussian uncertainty \cite{Muehlpfordt18c}. According to \cite{Warrington13}, any method including uncertainty should encompass three important aspects:
\begin{enumerate}
\itemsep0em 
	\item \label{item:forecasts} forecasts of uncertain disturbances such as feed-in from solar and wind sources, or demand;
	\item \label{item:policies} control policies for reacting to errors in forecasts;
	\item \label{item:propagation} propagation of uncertainties over time and space.
\end{enumerate}
Literature yet combines only aspects \ref{item:policies} and \ref{item:propagation}. Before we come to that, we list some applications for each aspect individually:
For the first aspect of forecasts of uncertain disturbances current literature proposes various methods that predict entire distributions or quantiles, see e.g. \cite{Hong16} for an overview, for different renewable energy sources \cite{quatile_zhang_2020, quantile_solar_2017}. 
For the second aspect of the control policies, affine control policies are often applied to problems related to the operation of power systems under uncertainty. These applications range from reserve planning \cite{Warrington13,Ding16,Bucher17} and dispatching in active distribution networks \cite{Fabietti17,Fabietti18}, to optimal power flow \cite{Vrakopoulou13b,Vrakopoulou17,Muehlpfordt18c,Louca16,MunozAlvarez14}, or building control \cite{Oldewurtel12}. 
Affine policies are already in use in power systems, and convince with their simple structure and their ability to split the power balance equation nicely such that it is fulfilled despite uncertainties~\cite{Muehlpfordt18c}.
For the third aspect of the propagation of uncertainty, efficient methods have been proposed. For example, scenario-based approaches \cite{Capitanescu12,Fabietti17,Vrakopoulou17}, and approaches employing polynomial chaos \cite{Muehlpfordt17a,Muehlpfordt16b,Muehlpfordt18c}. Other works study multi-period (propagation over time) \gls{opf} under uncertainty, but employ scenario-based or data-driven approaches~\cite{Vrakopoulou13,Vrakopoulou17b,Capitanescu12,Guo18}.

Approaches combining both affine policies and propagation over time and space are to be found in both robust and stochastic optimization. Robust optimization does not assume an underlying uncertainty distribution, hence, it cannot offer an exact reformulation. In stochastic optimization, on the other hand, there are several approaches. 
Several are multi-period \glspl{opf} with storage that assume Gaussian uncertainty, however, they often do not include CCs or affine policies. They use scenario-trees \cite{hemmati_stochastic_2016}, others look at AC power flow in a distribution network \cite{ayyagari_chance_2017}, or approximate the \glspl{cc} \cite{summers_stochastic_2015, li_analytical_2015}. While some works do offer an exact reformulation of CCs, they are either static \cite{bienstock_chance_2012}, lack storages \cite{bienstock_chance_2012, roald_analytical_2013}, or do not include affine policies \cite{sjodin_risk-mitigated_2012}.
Few approaches offer models including \glspl{cc} and a formulation into a \gls{socp}, but lack affine policies and time \cite{zhang_distributionally_2017}, look at the risk of cost functions without storage \cite{xie_distributionally_2018}, or apply different chance constraints \cite{Warrington13}. 
Most importantly, none of the existing approaches combines all three aspects using an exact reformulation of the whole problem such that the result is an equivalent formulation.
The latter approaches differ to the methodology introduced in the present paper, and often also in their problem objective. Also, many of them focus on detailed modeling of specific parts, while we hold our formulation general.

In the present paper, we therefore provide a computationally efficient and analytically exact model for optimal generator and storage schedules that combines all three aspects; forecasts, control policies and uncertainty propagation. Specifically, we optimize stochastic chance-constrained multi-period \gls{opf} for transmission systems that are subject to time-varying uncertainties and contain a large energy storage system. We choose to use Gaussian Processes (\gps) to describe the uncertain demand and generation, as they are well-suited to model power time series \cite{mitrentsis_probabilistic_2021}. \gps are very flexible~\cite{Roberts12} and allow a closed-form expressions of random variables. Since they consist of Gaussian distributions that stay Gaussian when passed through some linear operator (such as the linear DC OPF). This idea of "analytical reformulation" has been used in \cite{Li17}, only they focus on joint chance constraints. Several works have applied \gps to wind power forecasting~\cite{Kou13,Chen14}, solar power forecasting~\cite{Sheng18}, and electric load forecasting~\cite{Mori08,Leith04,Lloyd14,Rogers11,Blum13,McLoughlin13}. Given our modelling choice of \gps, the natural way to forecast uncertain disturbances for different time horizons is through Gaussian process regression (GPR) \cite{Rasmussen06} as it yields the desired mean and covariance for \gps. We then provide a tractable and exact reformulation of the \gls{opf} problem as a \gls{socp} by leveraging affine feedback policies, and by using the closed-form expressions for all occurring random variables. Additionally, we use different risk levels for the chance constraints -- not to be confused with the risk of the cost function \cite{hemmati_stochastic_2016}.

To the best of our knowledge there are no works that model a DC multi-period CC-OPF, with affine policies and Gaussian uncertainty, in a transmission network, that is reformulated into a tractable, analytically exact equivalent, convex \gls{socp} \emph{and} including forecast of uncertainties via Gaussian process regression. In contrast to most literature we extensively test our model on various network sizes from 5 to 300 nodes.

The remainder of the paper is structured as follows. \Cref{sec:modelling_all} states the grid, models uncertainties as Gaussian processes, and introduces affine policies. \Cref{sec:OPF} states the yet intractable optimization problem under uncertainty. \Cref{sec:SolutionMethodology} then reformulates the \opf problem as a tractable convex optimization problem, and comments on its computational characteristics. The case studies in \Cref{sec:CaseStudy} apply the proposed optimization approach to the \textsc{ieee} 5-bus, 39-bus, 57-bus, 118-bus, and 300-bus test cases and a complexity analysis is provided. Lastly, the results are discussed in \Cref{sec:discussion}.

\section{Modelling assumptions}
\label{sec:modelling_all}

The model of the electrical grid is at the core of the optimization. 
Let us consider a connected graph with $\nbus$ buses and $\nline$ lines 
under \dc power flow conditions 
for time instants $\mathcal{\horizon} = \{ 1, \dots, \horizon \}$.
Every bus $i \in \mathcal{\nbus} =  \{1, \dots, \nbus \}$ can contain a disturbance $\load_i(t), i\in\mathcal{\Load}\subseteq\mathcal{\nbus}$, (i.e., load or renewables), a thermal generation unit $\gen_i(t)$, $i\in\mathcal{\Gen}\subseteq\mathcal{\nbus}$, and a storage injection unit~$\storage_i(t)$, $i\in\mathcal{\Storage}\subseteq\mathcal{\nbus}$.

We denote the power excess/deficit at node $i$ and time $t$ as
\begin{equation}
    \pnet_i(t) = \load_i(t) + \gen_i(t) + \storage_i(t),
\end{equation}
which is also the total power export/influx into/from node $i$ needed to guard the nodal power balance \cite{horsch_linear_2017}. 

In the following we will model the uncertain disturbances, as well as generation and storage that react to the disturbance and are modelled accordingly.

\subsection{Uncertain Disturbances as Gaussian Processes}
\label{sec:GP}

Uncertain disturbances are loads and volatile feed-ins from renewable energies. We denote them by~$\load_i(t)$ at bus~$i \in \mathcal{\nbus}$ and time~$t \in \mathcal{\horizon}$.
Specifically, we assume in this paper that the uncertainty is Gaussian and that the disturbances have no spatial correlation, i.e. the random variables are independent. For wind in particular Gaussianity of the prediction error is reasonable through the central limit theorem, since a large set of agglomerated wind farms has a normally distributed power output \cite{hemmati_stochastic_2016}.  This uncertain disturbance
is the realization of a discrete-time stochastic process $\{ \rv{\load}_i(t) \: \forall t \in \mathcal{\horizon} \}$ given by\,\footnote{More precisely, $\{ \rv{\load}_i(t) \: \forall t \in \mathcal{\horizon} \}$ is only a snapshot of the overarching stochastic process $\{ \rv{\load}_i(t) \: \forall t \in \tilde{\mathcal{\horizon}} \}$, where $\tilde{\mathcal{\horizon}}$ is an infinite set, and $\mathcal{\horizon} \subset \tilde{\mathcal{\horizon}}$ a finite subset thereof.
We however neglect this subtlety for the sake of simplicity in the present paper.}
\begin{subequations}
	\label{eq:ConditionOnStochasticProcess}
	\begin{align}
	\label{eq:ConditionOnStochasticProcess_Mapping}
	\scalemath{0.92}{%
		\begin{bmatrix}
		\rv\load_i(1) \\
		\rv\load_i(2) \\
		\vdots \\
		\rv\load_i(\horizon) \\
		\end{bmatrix}
		= \underbrace{\begin{bmatrix}
			[\hat\load_i]_1\\
			[\hat\load_i]_2 \\
			\vdots \\
			[\hat\load_i]_{\horizon}
			\end{bmatrix}}_{=: \hat\load_i} +
		\underbrace{\begin{bmatrix}
			[\Load_i]_{11} & 0 & \hdots & 0 \\
			[\Load_i]_{21} & [\Load_i]_{22} & & 0 \\
			\vdots & & \ddots\\
			[\Load_i]_{\horizon 1} & & & [\Load_i]_{\horizon \horizon}]
			\end{bmatrix}}_{=: \Load_i} 
		\underbrace{\begin{bmatrix}
			[\rv\Xi_i]_1 \\
			[\rv\Xi_i]_2 \\
			\vdots \\
			[\rv\Xi_i]_\horizon		
			\end{bmatrix}}_{
			=: \rv\Xi_i}
	} 
	\end{align}
for all buses $i \in \mathcal{\nbus}$,
where $\hat\load_i \in \mathbb{R}^{\horizon}$ is the mean vector and $\Load_i \in \mathbb{R}^{\horizon \times \horizon}$ the lower-triangular, non-singular covariance matrix.
The stochastic germ~$\rv\Xi_i$ is a $\horizon$-variate Gaussian random vector whose elements are independent Gaussian random variables $[\rv\Xi_i]_j\sim\mathcal{N}(0,1)$.\,\footnote{Notice that non-singularity of $\Load_i$ means that~\eqref{eq:ConditionOnStochasticProcess_Mapping} is a one-to-one mapping between $[\rv\load_i(1), \dots,	\rv\load_i(\horizon)]^\top$ and the stochastic germ~$\rv\Xi_i$. The lower-triangularity of $\Load_i$ allows to create this mapping first for time instant $t=1$, then $t=2$, etc.\label{foot:OnetoOneMapping}}
Hence, the forecast error is Gaussian.
The lower-triangularity of $\Load_i$ means that the uncertain disturbance~$\rv\load_i(t)$ is causal, i.e. 
\begin{align}
\label{eq:UncertaintyModel}
\rv\load_i(t) = [ \hat\load_i ]_t + \sum_{k=1}^{t} [\Load_i]_{tk} [\rv\Xi_i]_k,
\end{align}
where $\rv\load_i(t)$ depends only on past and present time instants $k = 1, \dots, t$, but not on future ones.
\end{subequations}
Every uncertain disturbance is then fully described by its mean~$\ev{\rv\load_i(t)}$ and variance~$\var{\rv\load_i(t)}$, which we need to provide for the given time horizon
\begin{equation}
\label{eq:MeanVariance_Load}
\ev{\rv\load_i(t)} = [ \hat\load_i ]_t, \quad
\var{\rv\load_i(t)} = \sum_{k=1}^{t} [\Load_i]_{tk}^2.
\end{equation}

\subsection{Affine Policies}

Having parametrized the uncertain disturbances in an affine fashion, the reaction of generation and storage is modelled accordingly. In particular, the latter have to to assume uncertainty themselves as uncertainty means that they can react to errors in forecasts. Otherwise, the power balance equation could not be fulfilled. Therefore, we model generation and storage analogously to the uncertainty: as realizations of (affine) random processes $\{ \rv\gen_i(t) \, \forall t \in \mathcal{\horizon} \}$ and $\{ \rv\storage_i(t) \, \forall t \in \mathcal{\horizon} \}$, respectively.

We do that by introducing \textit{affine policies} that determine how generation and storage react to the uncertain disturbances. For generation we introduce feedback of the form
\begin{subequations}
\label{eq:GenerationPolicy}
\begin{align}
\label{eq:AffineFeedback_Gen}
\rv\gen_i
= \hat{\gen}_i + \sum_{j \in \mathcal{\nbus}} \Gen_{i,j} \rv\Xi_j, \quad \forall i \in \mathcal{\nbus},
\end{align}
\end{subequations}
for all time instants $t \in \mathcal{\horizon}$.\,\footnote{Notice that the control policy~\eqref{eq:AffineFeedback_Gen} is written in terms of the stochastic germs~$\rv\Xi_j$ for $j \in \mathcal{\nbus}$;
but in practice it is the realization of the uncertain disturbances $\rv\load_i(t)$ that can be measured.
It is always possible to get the realization of the stochastic germ from the realization of the uncertain disturbance, and vice versa, see Footnote~\ref{foot:OnetoOneMapping}.\label{foot:Rewriting}}
For this, $\hat{\gen}_i \in \mathbb{R}^{\horizon}$, and every $\Gen_{i,j} \in \mathbb{R}^{\horizon \times \horizon}$ with $j \in \mathcal{\nbus}$ is lower-triangular. 
The latter enforces the feedback to be causal, as they cannot depend on future uncertainties.
Note that the notation is structurally equivalent to~\eqref{eq:ConditionOnStochasticProcess} with the same stochastic germ.

We introduce the same kind of feedback policy~\eqref{eq:AffineFeedback_Gen} for the storage injections (from storage to grid)
\begin{align}
\label{eq:StoragePolicyPerBusPerTime}
\rv\storage_i
= \hat{\storage}_i + \sum_{j \in \mathcal{\nbus}} \Storage_{i,j} \rv\Xi_j,
\end{align}
where $\hat{\storage}_i\in\mathcal{R}^{\horizon}$ and every $\Storage_{i,j} \in \mathbb{R}^{\horizon \times \horizon}$ with $j \in \mathcal{\nbus}$ is lower-triangular.

Having established $\load_i(t)$, $\gen_i(t)$ and $\storage_i(t)$ we can further derive closed-form expressions of the other random variables.
From storage injections $\rv\storage_i(t)$ we can directly model the actual storage states $\rv\energy_i(t)$ as discrete-time integrators
\begin{align}
\label{eq:StorageDynamics_RV}
\rv\energy_i(t+1) = \rv\energy_i(t) - h \, \rv\storage_i(t), \quad \rv\energy_i(1) = \rv\energy_i^{\textsc{ic}} \quad \forall i \in \mathcal{\nbus}.
\end{align}
Reformulating the equation towards $\rv\storage_i(t)$ the denominator $h \, \rv\storage_i(t) = \rv\energy_i(t)-\rv\energy_i(t+1)$ makes clear that $\rv\storage_i(t)$ is the discharge of storage from time $t$ to $t+1$, or the injection into the network.
In general, uncertainty also affects the initial condition~$\rv\energy_i^{\textsc{ic}}$ of storage~$i$.
For simplicity, the value of $h > 0$ subsumes the discretization time and a potential loss factor. 

Moreover, the change of generation inputs can be derived as $\Delta\rv\gen_i(\tau) = \rv\gen_i(\tau) {-} \rv\gen_i(\tau {-} 1)$ and the net power becomes $\rv\pnet_i(t) = \rv\load_i(t) + \rv\gen_i(t) + \rv\storage_i(t)$ for bus $i$. 
Lastly, using the power transfer distribution matrix $\ptdfmat$ mapping net power to line flows, the line flow can be expressed as $\rv\lineflow_j(t) = \ptdfmat_{j} [\rv\pnet_1(t), \dots, \rv\pnet_{\nbus}(t)]^\top$. 
The voltage angles are implicitly contained in the definition of the net power $\rv\pnet_i(t)$ \cite{horsch_linear_2017}.
Note that all those random variables are Gaussian processes by linearity. Hence, as such they are fully described by their mean and variance, as listed in Table~\ref{tab:Moments}.

\begin{table*}
	\centering
	\caption{Closed-form expressions for state of storage~$\rv\energy_i(t+1)$, change of inputs~$\Delta\rv\gen_i(\tau)$, and line flows~$\rv\lineflow_j(t)$.\label{tab:FunctionalDepenendencies}}
	\small
	\begin{tabular}{lll}
		\toprule
		$ \rv\lineflow_l(t) $ & $=$ & $  \displaystyle\sum_{i \in \mathcal{\nbus}} [\ptdfmat]_{li} ( [\hat\load_i]_t + [ \hat\gen_i ]_t +[ \hat\storage_i ]_t ) + \displaystyle\sum_{i \in \mathcal{\nbus}} \sum_{k=1}^{t} \Big( [\ptdfmat]_{li} [\Load_i]_{tk} + \displaystyle \sum_{j \in \mathcal{\nbus}} [\ptdfmat]_{lj} ( [\Gen_{j,i}]_{tk} + [\Storage_{j,i}]_{tk} ) \Big) [\rv\Xi_i]_k $ \\
		$\Delta\rv\gen_i(\tau)$ & $=$ & $ [\hat\gen_i]_{\tau} - [\hat\gen_i]_{\tau {-} 1} + \displaystyle \sum_{j \in \mathcal{\nbus}} \Big( [\Gen_{i,j}]_{\tau \tau} [\rv\Xi_j]_{\tau} + \displaystyle \sum_{k=1}^{\tau {-} 1} \Big( [\Gen_{i,j}]_{\tau k} - [\Gen_{i,j}]_{(\tau {-} 1) k} \Big) [\rv\Xi_j]_k \Big)$ \\
		$ \rv\energy_i(t+1)$ & $=$ & $\rv\energy_i^{\textsc{ic}} - h \displaystyle \sum_{k=1}^t [\hat\storage_i]_k - h \displaystyle \sum_{j \in \mathcal{\nbus}} \sum_{k=1}^t \Big( \displaystyle \sum_{l=k}^t [\Storage_{i,j}]_{lk} \Big) [\rv\Xi_j]_k $ \\
		\bottomrule
	\end{tabular}
\end{table*}
\begin{table*}
	\centering
	\caption{Expected value and variance of random variables from Problem~\eqref{eq:CCOPF_original} under affine policies \eqref{eq:GenerationPolicy} and \eqref{eq:StoragePolicyPerBusPerTime}.\label{tab:Moments}}
	\small
	\begin{tabular}{lll|lll}
		\toprule
		$\rv x$ & $\ev{\rv x}$ & $\var{\rv x} = \sigma^2$ & $\rv x$ & $\ev{\rv x}$ & $\var{\rv x} = \sigma^2$\\
		\midrule
		$\rv\load_i(t)$ & $[\hat\load_i]_t$ & $\displaystyle\sum_{k=1}^{t} [\Load_i]_{tk}^2$ & $\rv\lineflow_l(t)$ & $\displaystyle\sum_{i \in \mathcal{\nbus}} [\ptdfmat]_{li} ( [\hat\load_i]_t + [ \hat\gen_i ]_t +[ \hat\storage_i ]_t )$ & $ \displaystyle\sum_{i \in \mathcal{\nbus}} \displaystyle \sum_{k=1}^{t} \Big(  [\ptdfmat]_{li} [\Load_i]_{tk} + \displaystyle \sum_{j \in \mathcal{\nbus}} [\ptdfmat]_{l,j} ([\Gen_{j,i}]_{tk} + ([\Storage_{j,i}]_{tk}) \Big)^2$\\
		$\rv\gen_i(t)$ & $[ \hat\gen_i ]_t$ & $\displaystyle\sum_{j \in \mathcal{\nbus}} \displaystyle\sum_{k=1}^{t} [ \Gen_{i,j} ]_{tk}^2$ & $\Delta\rv\gen_i(\tau)$ & $[ \hat\gen_i ]_{\tau} - [ \hat\gen_i ]_{\tau{-}1}$ &  $\displaystyle\sum_{i \in \mathcal{\nbus}} \Big( [\Gen_{i,j}]_{\tau \tau}^2 {+} \sum_{k=1}^{\tau{-}1} ( [\Gen_{i,j}]_{\tau k} - [\Gen_{i,j}]_{(\tau-1) k})^2 \Big) $\\
		$\rv\storage_i(t)$ & $[ \hat\storage_i ]_t$ & $\displaystyle\sum_{j \in \mathcal{\nbus}} \displaystyle\sum_{k=1}^{t} [ \Storage_{i,j} ]_{tk}^2$ & $\rv\energy_i(t+1)$ & $\ev{\rv\energy_i^{\textsc{ic}}} - h \displaystyle \sum_{k=1}^{t} [ \hat\storage_i ]_k $	 & $\var{\rv\energy_i^{\textsc{ic}}} + h^2 \displaystyle \sum_{j \in \mathcal{\nbus}} \displaystyle \sum_{k=1}^{t} \Big(\displaystyle \sum_{l=k}^{t} [\Storage_{i,j}]_{lk} \Big)^2$ \\	
		\bottomrule
	\end{tabular}
\end{table*}

\subsection{Local and global balancing}
\label{sec:localglobal}

We have formulated the generators response to uncertainty through affine policies. Furthermore, we can specify how exactly generators react through the structure of the matrices ~$\Gen_{i,j}$, called \emph{local} and \emph{global balancing}.

In local balancing each generator $i$ reacts to every possible source of uncertainty $\rv{\Xi}_j$
Global balancing lets each generator react to the \emph{sum} of deviations and can be achieved by enforcing $\Gen_{i,1} = \hdots = \Gen_{i,\nbus}$ \cite{Muehlpfordt18c}.


%

\subsection{Predicting Uncertainties with Gaussian Process Regression}
\label{sec:GPR}
To predict the uncertain disturbances~$\rv\load_i$, we need the mean $\hat\load_i$ and covariance matrix $\Load_i$.
Gaussian process regression (GPR) is a prediction method that yields precisely those.
GPR fits a family of functions $\mathcal{F}$ individually onto a data set $\mathcal{X}$. The posterior Gaussian process is then determined by the mean functions $\mu(t) = \mathbb{E}[\mathcal{F}(t)]$ ($t\in\mathbb{R}$) of $\mathcal{F}$ and a continuous covariance function $k(t,t')$
\footnote{$k$ is also called a \textit{kernel} and should be a positive definite function.}
, $t,t'\in\mathbb{R}$, yielding $\Load_i$.
Thereby, $k$ reflects both the variance around $\mu(t)$ for some $t$, as well as the covariance of two values for $t,t'$. We write the Gaussian process as $\mathcal{N}(\mu,k)$.
Since both $\mu$ and $k$ are continuous ($t\in\mathbb{R}$), for the prediction we can simply extract the discrete vector $\mu(t)\stackrel{\wedge}{=}\hat\load_i(t)$ and matrix $\Load_i$ by inserting all future $t\in\horizon$ into $\mu(t)$ and $(t,t')\in\horizon\times\horizon$ into $k$.
Then the Gaussian process at node $i$ is written as
\begin{equation}
    d_i = \mathcal{N}(\hat\load_i,\Load_i^2) \quad \forall i \in \nbus.
\end{equation}
For the kernel function $k$ we use the sum of cosine and squared exponential (i.e. RBF) with an added constant function--yielding
\begin{equation}
\label{eq:kernel}
    k = k_{cosine} + k_{RBF} + k_{constant},
\end{equation}
with
\begin{align*}
    &k_{cosine}(x,x') & &= \sigma_{1}^2\cos\left(2\pi \sum_i \frac{(x-x')}{l_{1}}\right), \\
    &k_{RBF}(x,x')    & &= \sigma_{2}^2 \exp{\left(-\frac{(x-x')^2}{2l_{2}^2}\right)}, \\
    &k_{constant}(x,x') & &= \sigma_{3},
\end{align*}
where $\sigma_i$ is the variance and $l_i$ the lengthscale parameter. The variance determines the average distance of some $f\in\mathcal{F}$ to the mean function $\mu=\mathbb{E}[\mathcal{F}(x)]$; the lengthscale determines the length of the 'wiggles' in $f$ \cite{duvenaud_automatic_nodate}. This allows us to model periodicity as well as larger trends and smaller variations.

Having modelled all decision variables as random variables (and described how the uncertain disturbance are obtained), we can now put them all together into an optimization problem.
\vspace{1em}


\section{Optimization problem for power systems under uncertainty}
\label{sec:OPF}

Given a network and Gaussian decision variables, we can now introduce constraints and an objective in order to formulate the optimal power flow problem. Besides limits for line flows, storage injections, states and final states, generators and change of generation, a main constraint is the power balance equation
\begin{equation}
    \sum_{i \in \mathcal{\nbus}} \rv\pnet_i(t) = 0.
\end{equation}
Note that this is not the nodal power balance equation as $\pnet$ is the excess/deficit at node $i$. 
The leading objective can be formulated as: \textit{''How can we operate generators optimally in the presence of uncertainty?''} (given storage systems) and we thus formulate the chance-constrained \opf problem as
\newcommand{\adjustCCOPF}{-10mm}
\begin{subequations} 
	\label{eq:CCOPF_original}
	\begin{align}
	\label{eq:CCOPF_original_cost}
	\underset{\rv{\gen}_i(t), \rv{\storage}_i(t)}{\operatorname{min}}~  & \sum_{t \in \mathcal{\horizon}}\sum_{i \in \mathcal{\nbus}} \ev{ f_i(\rv\gen_i(t)} \quad \mathrm{s.t.}\\
	\label{eq:CCOPF_original_PowerBalance}
	& \hspace{\adjustCCOPF} \sum_{i \in \mathcal{\nbus}} \rv\load_i(t) + \rv\gen_i(t) + \rv\storage_i(t) = 0 \\
	\label{eq:CCOPF_original_StorageDynamics}
	& \hspace{\adjustCCOPF} 	\rv\energy_i(t+1) = \rv\energy_i(t) - h \, \rv\storage_i(t), ~ \rv\energy_i(1) = \rv\energy_i^{\textsc{ic}} \\
	\label{eq:CCOPF_CCs}
	& \hspace{\adjustCCOPF} \prob{ \rv x(t) \leq \overline{x} } \geq 1 - \varepsilon,
	~ \prob{ \rv x (t) \geq \underline{x} } \geq 1 - \varepsilon  \\
	\label{eq:CCOPF_original_StdConstraint}
	& \hspace{\adjustCCOPF} 0 \leq \sqrt{\var{\rv{x}}} \leq \sigma_{\overline{\rv{x}}} \\
	& \hspace{\adjustCCOPF} \forall \rv{x} {\in} \{ \rv\lineflow_j (t), \rv\gen_i(t), 
	\Delta\rv\gen_i(\tau),\! \rv\energy_i(t {+} 1), \rv\energy_i(T), \rv\storage_i(t) \} \\
	\nonumber
	& \hspace{\adjustCCOPF} \forall i \in \mathcal{\nbus}, \, t \in \mathcal{\horizon}, \, \tau \in \mathcal{\horizon} \setminus \{1\}, j \in \mathcal{L},
	\end{align}
\end{subequations}
where $\varepsilon \in (0,0.1]$ is the risk factor.\,\footnote{It is straightforward to modify Problem~\eqref{eq:CCOPF_original} to consider time-varying and quantity-depending risk levels $\varepsilon$, e.g. use $\overline\varepsilon_{\lineflow_j}(t)$ to specify the risk level for satisfying the upper limit of line $j$ at time $t$.}
Problem~\eqref{eq:CCOPF_original} minimizes the expected cost of generation over time~\eqref{eq:CCOPF_original_cost},
while satisfying the power balance~\eqref{eq:CCOPF_original_PowerBalance}
and the storage dynamics~\eqref{eq:CCOPF_original_StorageDynamics} in terms of random processes. 
\footnote{For ease of presentation we assume the storage has already been installed and that their operation does not incur costs.}

All engineering limits are formulated with chance constraints~\eqref{eq:CCOPF_CCs}: the probability that the line flow $\rv\lineflow_j(t)$, the generation $\rv\gen_i(t)$, the generation ramp $\Delta\rv\gen_i(\tau)$, the storage $\rv\storage_i(t)$, $\rv\energy_i(t)$ are below/above their upper/lower limits shall be greater than or equal to $1 - \varepsilon$.
We add chance constraints for the terminal state of the storage, $\rv\energy_i(T)$, to allow for the storage to be at a predefined level (with high probability) at the end of the horizon. The inequality constraint~\eqref{eq:CCOPF_original_StdConstraint} allows to restrict the standard deviation of all occurring random variables. The restriction enables to reduce the variation of certain generation units to be small. Note that this model can easily be adapted to power plants without ramp constraints (e.g. gas plants), by removing the respective equations.

\begin{figure}
    \centering
    \includegraphics[width=0.55\textwidth]{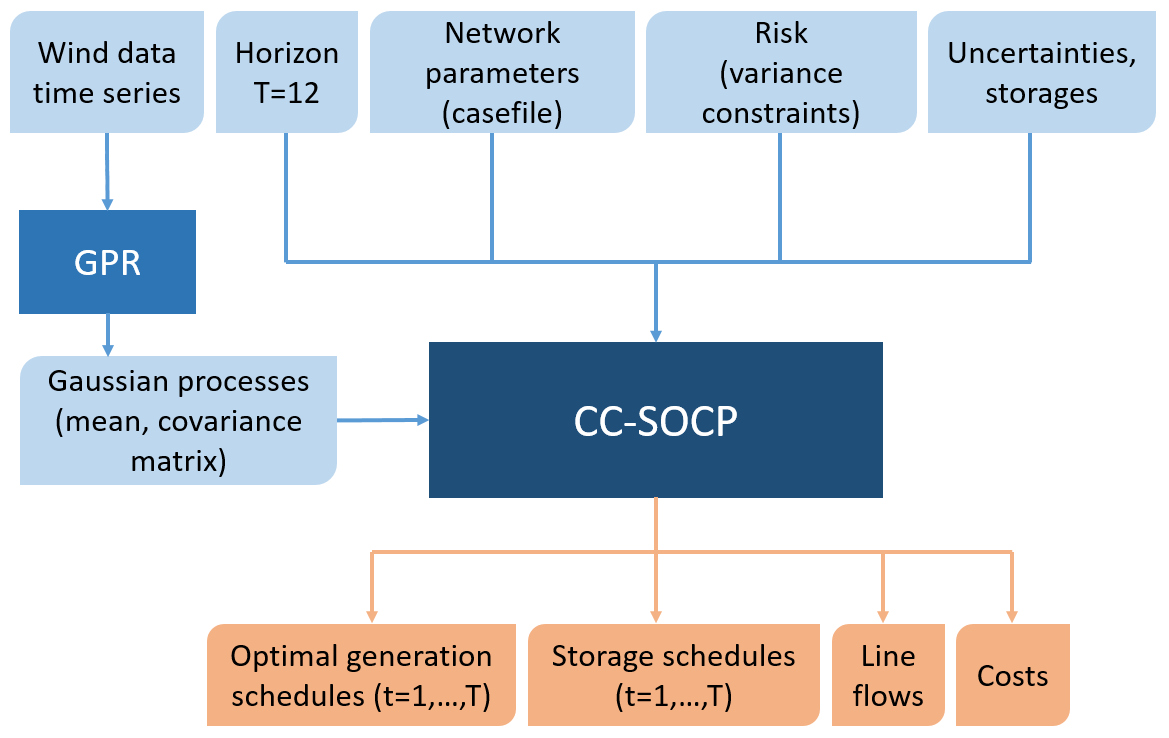}
    \caption{Inputs and outputs of the dynamic CC-SOCP.}
    \label{fig:model_input_output}
\end{figure}

Figure \ref{fig:model_input_output} visualizes this method, where the inputs are network parameters, uncertainties and storage, the time horizon, risk parameter, and predicted wind power as Gaussian processes. The outputs are then the optimal generation (decision variable) and its costs (objective), as well as storage schedules and line flows.


\section{Reformulation of Optimization Problem}
\label{sec:SolutionMethodology}

Problem~\eqref{eq:CCOPF_original} is intractable for several reasons: the decision variables are random processes, the equality constraints are infinite-dimensional, and the chance constraints and cost function require to evaluate integrals for the chance-constraints. In order to derive an exact yet finite-dimensional reformulation of the problem and cope with the intractability issues, we exploit the problems structure and the Gaussianity of all random variables. More specifically, we reformulate the infinite-dimensional power flow equation, compute the probabilities of the chance constraints, and rephrase the cost function.

\subsection{Power Balance}
\label{sec:PowerBalance}

To adapt the optimal power flow equations we start by taking the power balance~\eqref{eq:CCOPF_original_PowerBalance} and substituting both the uncertainty model~\eqref{eq:ConditionOnStochasticProcess} and the generation/storage control policies~\eqref{eq:GenerationPolicy}. Then, the power balance is satisfied for all realizations if \cite{Muehlpfordt18c}
\begin{subequations}
	\label{eq:PowerBalance_reformulated}
	\begin{align}
	\label{eq:PowerBalance_reformulated_mean}	
	\sum_{i \in \mathcal{\nbus}} \hat\load_i + \hat\gen_i + \hat\storage_i &= 0_{\horizon}, \\
	\label{eq:PowerBalance_reformulated_var}	
	\Load_j + \sum_{i \in \mathcal{\nbus}} \Gen_{i,j} + \Storage_{i,j} &= 0_{\horizon \times \horizon}, & \forall j \in \mathcal{\nbus}.
	\end{align}
\end{subequations}
Equation~\eqref{eq:PowerBalance_reformulated_mean} ensures power balance in the absence of uncertainties, or equivalently power balance in terms of the expected value; equation~\eqref{eq:PowerBalance_reformulated_var} balances every uncertainty~$\Load_j$ by the sum of the reactions from generation and storage.

\subsection{Chance Constraints}
\label{sec:ChanceConstraints}
As all random variables occurring in Problem~\eqref{eq:CCOPF_original} are Gaussian random variables, the chance constraints can be reformulated exactly using the first two moments:
Let $\rv x$ be a Gaussian random variable with mean $\mu$ and variance $\sigma^2$.
Then for $\varepsilon \in (0,0.1]$,
\begin{subequations}
	\label{eq:CC_Reformulation}
\begin{align}
&\prob{ \rv x \leq \overline{x}} \geq 1 - \varepsilon \quad \Longleftrightarrow & \hspace{-0.6cm} \phantom{\underline{x} \leq }\,\,\mu + \lambda(\varepsilon) \sqrt{\sigma^2} \leq \overline{x}, \\
&\prob{ \underline{x} \leq \rv x } \geq 1 - \varepsilon \quad  \Longleftrightarrow & \hspace{-0.6cm} \underline{x} \leq \mu - \lambda(\varepsilon) \sqrt{\sigma^2},
\end{align}
\end{subequations}
where $\lambda(\varepsilon) = \Psi^{-1}(1 {-} \varepsilon)$, and $\Psi$ is the cumulative distribution function of a standard Gaussian random variable \cite{Bienstock14}.
Hence, all chance constraints from Problem~\eqref{eq:CCOPF_original} can be reformulated by applying relation~\eqref{eq:CC_Reformulation} with the moments from Table~\ref{tab:Moments}.
Similarly, the constraint on the standard deviation~\eqref{eq:CCOPF_original_StdConstraint} is rewritten exactly using the expressions from Table~\ref{tab:Moments}.

\subsection{Cost Function}
\label{sec:CostFunction}
To rephrase the cost function, we consider quadratic generation costs
\begin{subequations}
\label{eq:Cost_Reformulation}
\begin{align}
f_i(\rv\gen_i(t)) = \gamma_{i,2} \rv\gen_i(t)^2 + \gamma_{i,1} \rv\gen_i(t) + \gamma_{i,0},
\end{align}
with $\gamma_{i,2} > 0$ for all buses $i \in \mathcal{\nbus}$. However, for a tractable problem we need scalar values in the objective function, not stochastic variables. A common technique is to simply take the expected value. This leads to the new objective function
\begin{align}
\ev{f_i(\rv\gen_i(t))} =  f_i(\ev{\rv\gen_i(t)}) +  \gamma_{i,2}\var{\rv\gen_i(t)} .
\end{align}
\end{subequations}

\subsection{Second-Order Cone Program}

Finally, by combining the results from Sections~\ref{sec:PowerBalance}, \ref{sec:ChanceConstraints}, and \ref{sec:CostFunction}, we present a finite-dimensional and tractable reformulation of Problem~\eqref{eq:CCOPF_original}:
\begin{subequations} 
	\label{eq:SOCP}
	\begin{align}
	\underset{\substack{\hat\gen_i,\, \Gen_{i,j}, \\ \hat\storage_i,\, \Storage_{i,j} \\
	\forall i,j \in \mathcal{\nbus}}}{\operatorname{min}}~  & \sum_{t \in \mathcal{\horizon}}\sum_{i \in \mathcal{\nbus}} f_i(\ev{\rv\gen_i(t)}) +  \gamma_{i,2}\var{\rv\gen_i(t)}  \quad \mathrm{s.\,t.} \\
	\begin{split}
		&\sum_{i \in \mathcal{\nbus}} \hat\load_i + \hat\gen_i + \hat\storage_i = 0_{\horizon}\\
		&\Load_j + \sum_{i \in \mathcal{\nbus}} \Gen_{i,j} + \Storage_{i,j} = 0_{\horizon \times \horizon}, \quad  \forall j \in \mathcal{\nbus}
	\end{split} \\
	& \rv\energy_i(t+1) = \text{\{see Table \ref{tab:FunctionalDepenendencies}\}}, \quad \rv\energy_i(1) = \rv\energy_i^{\text{\textsc{ic}}} \\
	& \underline{x} \leq \ev{\rv{x}} \pm \lambda(\varepsilon) \sqrt{\var{\rv{x}}} \leq \overline{x} \\
	& \sqrt{\var{\rv{x}}} \leq \overline{x}_{\sigma} \\
	\nonumber
	& \forall \rv{x} \in \{ \rv\lineflow_j(t), \, \rv\gen_i(t), \, \Delta\rv\gen_i(\tau), \, \rv\energy_i(t+1), \,\rv\energy_i(\horizon),\, \rv\storage_i(t) \} \\
	\nonumber
	& \forall i \in \mathcal{\nbus}, \, t \in \mathcal{\horizon}, \, \tau \in \mathcal{\horizon} \setminus \{1\}, j \in \mathcal{L}.
	\end{align}
\end{subequations}
Problem~\eqref{eq:SOCP} is a second-order cone program (\socp), hence a convex optimization problem. 

Let us add two more notes on the exact solution and number of decision variables:
As a first note, the \socp provides an \emph{exact} reformulation of Problem~\eqref{eq:CCOPF_original} in the following sense: let $(\rv{\gen}_i(t)^\star, \rv\storage_i(t)^\star)$ for all $i \in \mathcal{N}$ denote the optimal solution to Problem~\eqref{eq:CCOPF_original} restricted to the affine policy~\eqref{eq:AffineFeedback_Gen}, and
let $(\hat{\gen}_i^\star, \Gen_{i,j}^\star, \hat{\storage}_i^\star, \Storage_{i,j}^\star)$ for all $i, j \in \mathcal{\nbus}$ denote the optimal solution to \socp~\eqref{eq:SOCP}.
Applying~\eqref{eq:CC_Reformulation} and~\cite[Proposition 1]{Muehlpfordt17a}, the optimal policies for Problem~\eqref{eq:CCOPF_original} are given by the optimal values of the policy \emph{parameters} via Problem~\eqref{eq:SOCP}
\begin{align}
\begin{bmatrix}
\rv\gen_i(t)^\star \\
\rv\storage_i(t)^\star
\end{bmatrix}
= 
\begin{bmatrix}
[ \hat\gen_i^\star ]_t \\
[ \hat\storage_i^\star ]_t
\end{bmatrix}
+
\sum_{j \in \mathcal{\nbus}} \sum_{k=1}^{t}
\begin{bmatrix}
[   \Gen_{i,j}^\star ]_{tk}   
\\
[   \Storage_{i,j}^\star ]_{tk}
\end{bmatrix}
[\rv\Xi_j]_k
\end{align}
for all buses $i \in \mathcal{\nbus}$ and time instants $t \in \mathcal{\horizon}$.

A second note is that, in theory, the problem is tractable and should be solved efficiently with certified optimality in case of a zero duality gap. However, in practice, large grids may be numerically challenging due to many uncertainties and long horizons $\horizon$. Therefore, it is advisable to introduce a minimum number of scalar decision variables. Specifically, assuming that no bus has both a generator \emph{and} storage, i.e. $\mathcal{\Gen} \cap \mathcal{\Storage} = \emptyset$, for a grid with $N_{\load}$ disturbances, $N_{\gen}$ generators, and $N_{\storage}$ storage systems sets the number of decision variables for local balancing to
\begin{equation}
\label{eq:DecisionVariables_local}
(N_\gen + N_\storage) \left( \horizon + N_\load \frac{\horizon (\horizon + 1)}{2} \right),
\end{equation}
for the generation/storage policies~\eqref{eq:GenerationPolicy}/\eqref{eq:StoragePolicyPerBusPerTime}
\footnote{In contrast to \cite{Warrington13}, we exploit lower-triangularity of the matrices $\Gen_{i,j}$, $\Storage_{i,j}$.}
in local balancing.

In global balancing, see subsection~\ref{sec:localglobal}, for both generation and storage the number of scalar decision variables reduces to
\begin{equation}
\label{eq:DecisionVariables_global}
(N_\gen + N_\storage) \left( \horizon + \frac{\horizon (\horizon + 1)}{2} \right),
\end{equation}
hence it is independent of the number of uncertainties in the grid.
The difference between the numbers~\eqref{eq:DecisionVariables_local} and~\eqref{eq:DecisionVariables_global} reflects the usual trade-off between computational tractability and complexity of the solution.

To summarize: by using affine control policies the infinite-dimensional Problem~\eqref{eq:CCOPF_original} can be written as a tractable convex optimization problem. Since all reformulations are equivalent transformations, there is no loss of information, e.g. all chance constraints from Problem~\eqref{eq:CCOPF_original} are satisfied \emph{exactly}; there is no additional conservatism. Table \ref{tab:comparison_CCOPF_SOCP} illustrates this process.

\begin{table}[ht]
\centering
\caption{Comparison of Problem \eqref{eq:CCOPF_original} and \eqref{eq:SOCP}.}
\label{tab:comparison_CCOPF_SOCP}
\begin{tabular}[t]{lcc}
\toprule
\footnotesize
& Formulation \eqref{eq:CCOPF_original} & Reformulation \eqref{eq:SOCP} \\
\midrule
Problem type    & No SOCP       & SOCP \\
\# constraints  & Infinite      & Finite \\
Solve CCs       & Integral      & Exact formulation \\
Variables       & Random process & Gaussian process \\
Convexity       & Not convex    & Convex \\
Tractability     & \textbf{No}    & \textbf{Yes} \\
\bottomrule
\end{tabular}
\end{table}%
\normalsize

\section{Case Studies}
\label{sec:CaseStudy}

We test the reformulated OPF on various standard test grids of different size. We start with examining a small network with 5 nodes (\textsc{ieee} case5) in Section \ref{sec:case5} as the solutions are easy to verify and understand. To show that the model works equally well on larger grids, we test the OPF on the 39-bus \textsc{ieee} test case in Section \ref{sec:case39}.
Finally, in Section \ref{sec:complexityAnalysis}, we perform a complexity analysis regarding computation time with the additional grids \textsc{ieee} case57, case118 and case300. 

For all networks, we test three scenarios; without storage (\caseNoStorage), with storage (\caseStorage) and with storage and variance constraints (\caseStorageWithVariance). The variance constraints are introduced by
\begin{align}
\label{eq:VarianceConstraint}
\sqrt{\var{\rv\gen_i(t)}} \leq 0.01.
\end{align}
We test different uncertain disturbances and storage sets, and compare local and global balancing. 
If not stated otherwise, the risk level for each chance constraint in Problem~\eqref{eq:CCOPF_original} is set to $\varepsilon = 5\,\%$ and local balancing is used. In the complexity analysis we use more risk levels ($\epsilon\in\{2.5\%, 5\%, 10\%\}$). 
There are no costs for storage usage; generation costs are the same for all generators. Additionally, storage systems have a prescribed final state, see constraints~\eqref{eq:CCOPF_CCs}, and a maximum capacity.

Apart from showing that the method works well, we answer (i) what importance storage has in a power system with uncertainty, (ii) how scalable our method is in terms of the number of uncertainties and storage, (iii) what influence variance constraints have, (iV) how local and global balancing differ, and (v) what influence different risk levels have.

For the wind forecasts we use a real world wind power data set from ENTSO-E \cite{de_felice_matteo_2021_4682697} that encompasses time series from 2014 to 2021. We smooth the time series with a rolling window of 10 hours and scale according to network capacities. Since the wind farms and data windows are chosen randomly, there is no spatial or temporal correlation that should be considered.

For the sake of simplicity, and without loss of generality, we use the following function to model loads with horizon~$t \in\mathcal{\horizon} = \{1,\dots,12\}$, and, for better understanding, we also use it as a simple, additional forecast for case5:
\begin{subequations}
\label{eq:artificial_forecast}
	\begin{align}
	- [\hat{\load}_i]_t & =  \load_{i}^{\text{nom}} (1 + 0.1 \sin(2 \pi (t-1)/\horizon)), \quad \forall i \in \mathcal{\nbus},\\
	- \Load_i &= 
	\begin{cases}
	\tilde\Load_i \text{ from \eqref{eq:Results_GPVariance_matrix}}, & \forall i \in \mathcal{\Load},\\
	0_{\horizon \times \horizon}, & \forall i \in \mathcal{\nbus} \cap \mathcal{\Load},
	\end{cases}
	\end{align}
\end{subequations}
where~$\load_{i}^{\text{nom}}$ is the nominal load value taken from the case files and $\tilde\Load_i$ is given by \eqref{eq:Results_GPVariance_matrix}.

\begin{figure}
	\centering
	\begin{equation}
	\label{eq:Results_GPVariance_matrix}
	\tilde\Load_i = 10^{-4}\cdot
	\scalemath{0.525}{
		\left[
		\begin{array}{cccccccccccc}
		87   & 0    & 0   & 0   & 0   & 0   & 0   & 0  & 0  & 0  & 0 & 0 \\
		176  & 20   & 0   & 0   & 0   & 0   & 0   & 0  & 0  & 0  & 0 & 0 \\
		292  & 60   & 7   & 0   & 0   & 0   & 0   & 0  & 0  & 0  & 0 & 0 \\
		434  & 124  & 26  & 3   & 0   & 0   & 0   & 0  & 0  & 0  & 0 & 0 \\
		594  & 211  & 63  & 13  & 3   & 0   & 0   & 0  & 0  & 0  & 0 & 0 \\
		764  & 321  & 123 & 31  & 13  & 3   & 0   & 0  & 0  & 0  & 0 & 0 \\
		937  & 447  & 208 & 63  & 32  & 11  & 3   & 0  & 0  & 0  & 0 & 0 \\
		1103 & 582  & 317 & 109 & 65  & 27  & 10  & 3  & 0  & 0  & 0 & 0 \\
		1257 & 718  & 447 & 172 & 116 & 55  & 26  & 10 & 3  & 0  & 0 & 0 \\
		1392 & 847  & 591 & 251 & 184 & 98  & 53  & 26 & 10 & 3  & 0 & 0 \\
		1504 & 964  & 741 & 342 & 271 & 156 & 94  & 53 & 24 & 9  & 3 & 0 \\
		1590 & 1063 & 889 & 441 & 371 & 229 & 151 & 94 & 50 & 24 & 9 & 3
		\end{array}
		\right]
	}
	\end{equation}
	\vspace{\adjustlength}
\end{figure}

For the Gaussian process regression we need to perform a Cholesky decomposition $\Load_i$ of the covariance matrix, to which we apply whitening of $1e^{-7}$ due to slight numerical instabilities.
Gaussian process regression was implemented in Python \cite{10.5555/1593511} version 3.8.8 using GpFlow \cite{GPflow2017} based on tensorflow.
The SOCPs were implemented in Julia~\cite{Bezanson2017} version 1.6.1, and solved with \textsc{j}u\textsc{mp}~\cite{Dunning2017} and the MOSEK solver set to its default values, using a PC with an AMD Ryzen™ 7 PRO 4750U processor at 1700 Mhz and 16GB memory \cite{muhlpfordt_git_nodate}.

\subsection{\textsc{ieee} 5-bus test case}
\label{sec:case5}

\begin{figure}
	\begin{subfigure}[c]{\figwidth}
    \centering
    \includegraphics[width=0.95\textwidth]{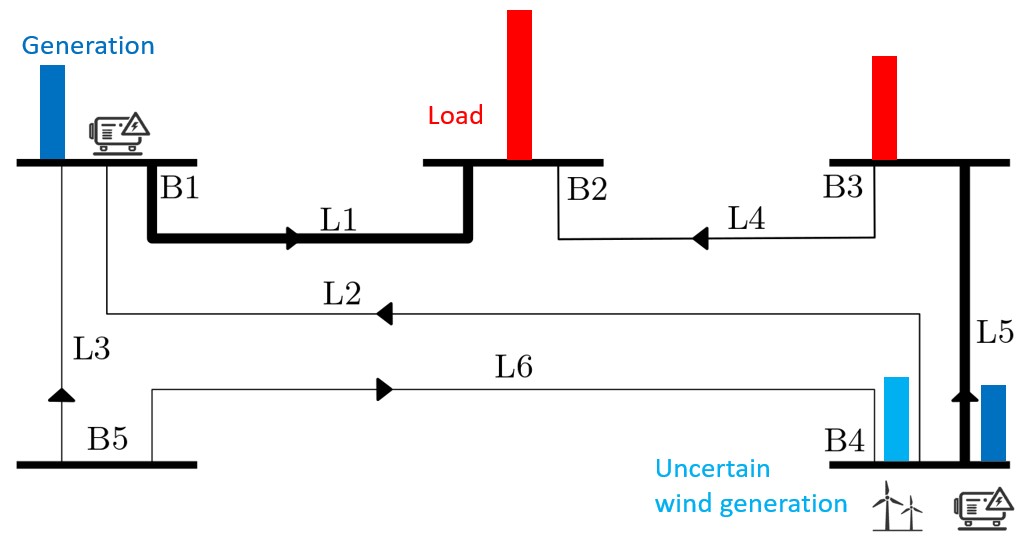}
    \caption{Without storage.}
    \label{fig:case5_results_grids_NoStorage}
\end{subfigure}
\begin{subfigure}[c]{\figwidth}
    \centering
    \includegraphics[width=0.95\textwidth]{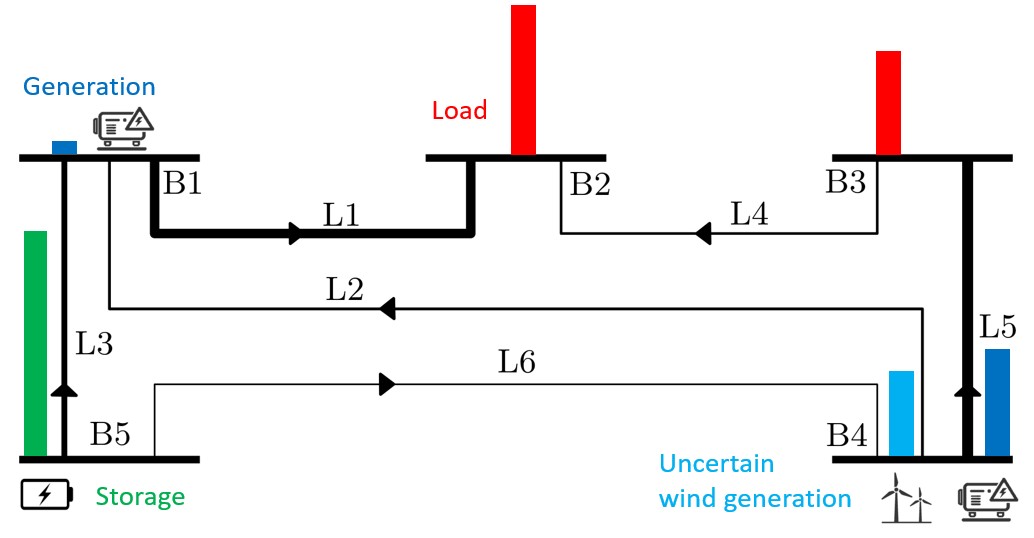}
    \caption{With storage.}
    \label{fig:case5_results_grids_Storage}
    \end{subfigure}
    \caption{\textsc{ieee} case5: Network state without (upper) and with (lower) storage at time $t=3$, with generation (dark blue), wind generation (light blue), loads (red), storage (green) and line flows (black).}
    \label{fig:case5:results_grids}
\end{figure}

Let us first apply method \eqref{eq:SOCP} to a simple test network in order to foster a good understanding of the dynamics. 
\textsc{ieee} case5 has five nodes, six lines, and two generators at buses $\mathcal{\Gen} = \{1,4\}$. We install two loads at buses $\{2,3\}$, one storage at bus $\mathcal{\Storage} = \{5\}$ and one uncertain disturbance at bus $\mathcal{\Load} = \{4\}$ that represents a wind farm, see Figure \ref{fig:case5:results_grids}. 

We alter the case file slightly in order to make it compatible with our method: Generators 1 and 2 are merged (by adding their capacities Pg, Qg, Qmax, Qmin, Pmax), because the program requires maximal one generator per node. 
And generator 5 is replaced by a storage, as each node can only contain a generator \emph{or} a storage. 
All minor changes, such as cost coefficients and line ratings, can be found in Table \ref{tab:Parameters}.

\begin{figure}
    \centering
    \includegraphics[width=0.5\textwidth]{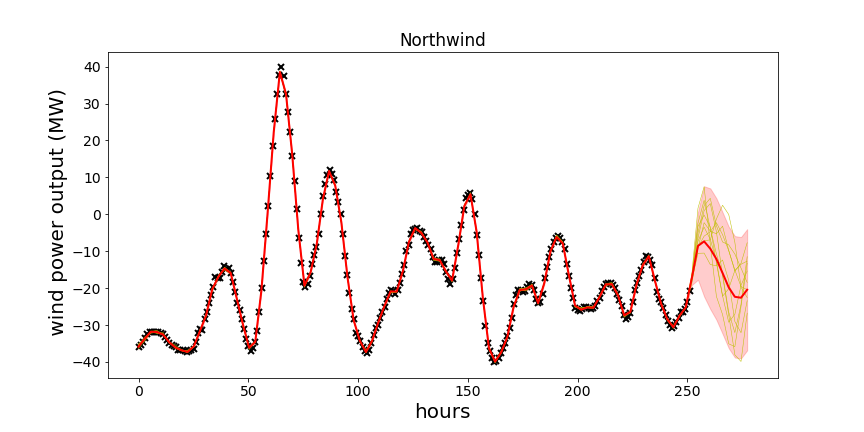}
    \caption{GPR-fitted and forecast wind power outputs smoothed with a rolling window of 5.}
    \label{fig:wind_forecast_volatile}
\end{figure}

Besides the network, the OPF requires a second input; wind forecast in the form of Gaussian processes.
Figure \ref{fig:wind_forecast_volatile} shows the forecast of wind power for a random day of wind farm \emph{Northwind}. We selected the kernel as in equation \eqref{eq:kernel}. 
As we can see, the GPR fits the given data well, while the horizon encompasses more variance (uncertainty).

\begin{figure}
    \begin{subfigure}[c]{\figwidth}
        \centering
        \includegraphics[width=0.45\textwidth]{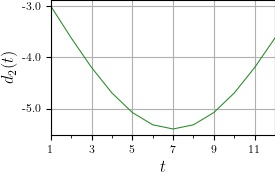}
        \includegraphics[width=0.45\textwidth]{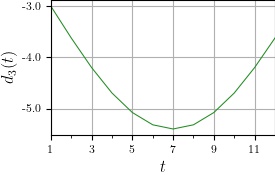}
        \includegraphics[width=0.6\textwidth, height=0.3\textwidth]{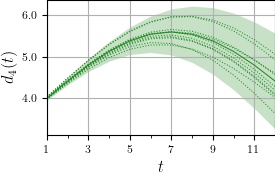}
        \vspace{-2mm}
        \caption{Upper: certain disturbances~$-\load_{2}(t), \load_{3}(t)$ at buses $\{2,3\}$; lower: ten realizations of the uncertain disturbance~$-\rv\load_4(t)$ at bus~$4$ (dots), mean $-\ev{\rv\load_4(t)}$ (solid), and $-(\ev{\rv\load_4(t)} \pm 3\sqrt{\var{\rv\load_4(t)}})$-interval (shaded).}
        \label{fig:case5:Loads}
    \end{subfigure}
    
	\begin{subfigure}[c]{\figwidth}
		\centering
		
        \includegraphics[width=0.45\textwidth]{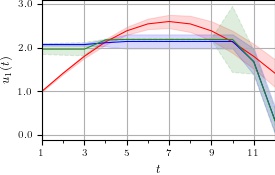}~
        \includegraphics[width=0.45\textwidth]{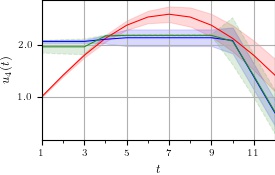}%
		
        \includegraphics[width=0.45\textwidth]{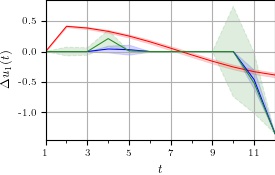}~
        \includegraphics[width=0.45\textwidth]{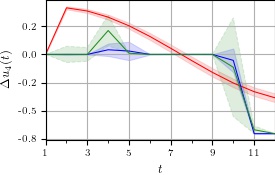}%
        
		\vspace{-2mm}		
		\caption{Upper: Power injections of generator at buses $\{1, 4\}$; lower: respective change in power injections.}
		\label{fig:case5:Generation}
	\end{subfigure}
	
	\begin{subfigure}[c]{\figwidth}
		\centering
        \includegraphics[width=0.45\textwidth]{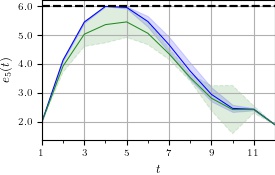}~
        \includegraphics[width=0.45\textwidth]{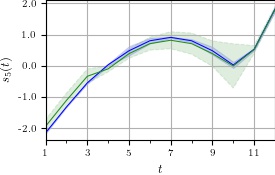}
		
		\vspace{-2mm}
		\caption{Left: Power injections of storage at bus $5$; right: respective change of power.}
		\label{fig:case5:Storage}
	\end{subfigure}
	\begin{subfigure}[c]{\figwidth}
	\centering
        \includegraphics[width=0.45\textwidth]{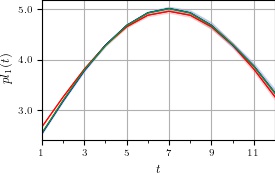}
        \includegraphics[width=0.45\textwidth]{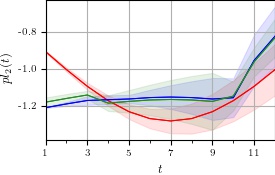}
        \includegraphics[width=0.45\textwidth]{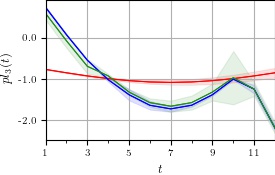}
        \includegraphics[width=0.45\textwidth]{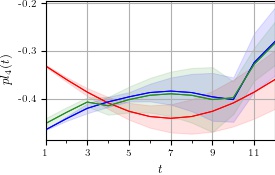}
        \includegraphics[width=0.45\textwidth]{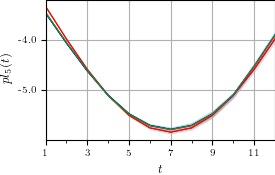}
        \includegraphics[width=0.45\textwidth]{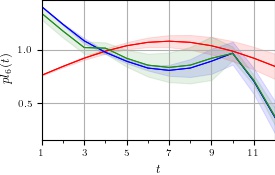}
		\vspace{-2mm}		
		\caption{Line flows across all lines.}
		\label{fig:case5:LineFlows}
	\end{subfigure}
	\vspace{\adjustlength}
	\caption{\textsc{ieee} 5-bus grid: Results for cases \caseNoStorage (red), \caseStorage (blue), and \caseStorageWithVariance (green). All shown random variables~$\rv{x}$ are depicted in terms of their mean~$\ev{\rv{x}}$ (solid) and the interval $\ev{\rv{x}} \pm \lambda (0.05) \sqrt{\var{\rv{x}}}$ (shaded).}
	\label{fig:case5:results_artificial_newLoads}
\end{figure}

\begin{figure}
    \begin{subfigure}[c]{\figwidth}
        \centering
        \includegraphics[width=0.45\textwidth]{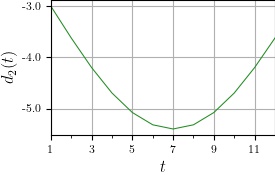}
        \includegraphics[width=0.45\textwidth]{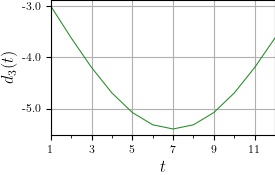}
        \includegraphics[width=0.6\textwidth, height=0.3\textwidth]{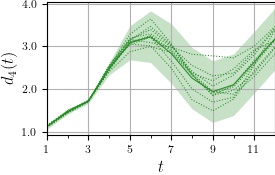}
        \vspace{-2mm}
        \caption{Upper: certain disturbances~$-\load_{2}(t), \load_{3}(t)$ at buses $\{2,3\}$; lower: ten realizations of the uncertain disturbance~$-\rv\load_4(t)$ at bus~$4$ (dots), mean $-\ev{\rv\load_4(t)}$ (solid), and $-(\ev{\rv\load_4(t)} \pm 3\sqrt{\var{\rv\load_4(t)}})$-interval (shaded).}
        \label{fig:case5:Loads_volatile}
    \end{subfigure}
	\begin{subfigure}[c]{\figwidth}
		\centering
        \includegraphics[width=0.45\textwidth]{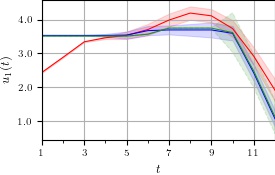}~
        \includegraphics[width=0.45\textwidth]{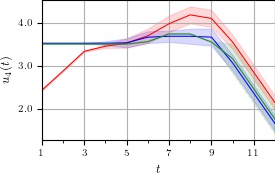}%
		
        \includegraphics[width=0.45\textwidth]{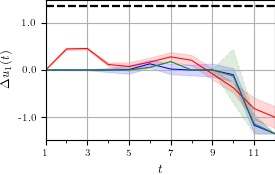}~
        \includegraphics[width=0.45\textwidth]{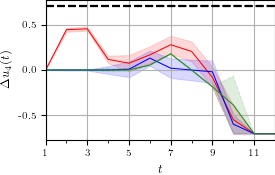}%
		\vspace{-2mm}		
		\caption{Upper: Power injections of generator at buses $\{1, 4\}$; lower: respective change in power injections.}
		\label{fig:case5:Generation_volatile}
	\end{subfigure}
	
	\begin{subfigure}[c]{\figwidth}
		\centering
        \includegraphics[width=0.45\textwidth]{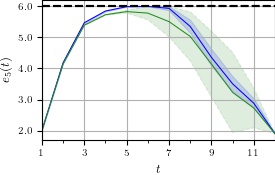}~
        \includegraphics[width=0.45\textwidth]{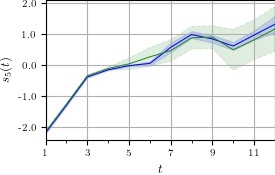}
		
		\vspace{-2mm}
		\caption{Left: Power injections of storage at bus $5$; right: respective change of power.}
		\label{fig:case5:Storage_volatile}
	\end{subfigure}
	\begin{subfigure}[c]{\figwidth}
	\centering
        \includegraphics[width=0.45\textwidth]{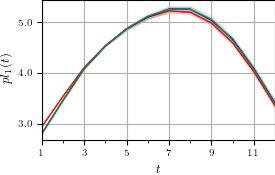}
        \includegraphics[width=0.45\textwidth]{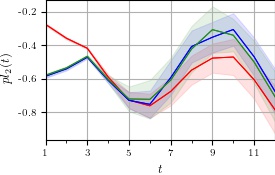}
        \includegraphics[width=0.45\textwidth]{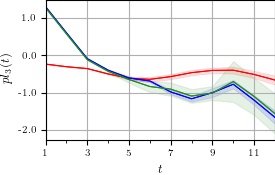}
        \includegraphics[width=0.45\textwidth]{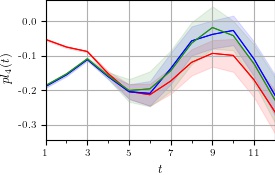}
        \includegraphics[width=0.45\textwidth]{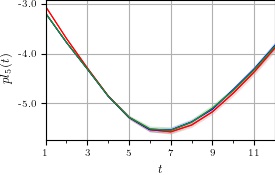}
        \includegraphics[width=0.45\textwidth]{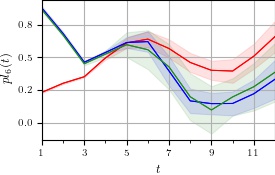}
		\vspace{-2mm}		
		\caption{Line flows across all lines.}
		\label{fig:case5:LineFlows_volatile}
	\end{subfigure}
	\vspace{\adjustlength}
	\caption{\textsc{ieee} 5-bus grid: Results for cases \caseNoStorage (red), \caseStorage (blue), and \caseStorageWithVariance (green). All shown random variables~$\rv{x}$ are depicted in terms of their mean~$\ev{\rv{x}}$ (solid) and the interval $\ev{\rv{x}} \pm \lambda (0.05) \sqrt{\var{\rv{x}}}$ (shaded).}
	\label{fig:case5:results_volatile_newLoads}
\end{figure}

The OPF results for the predicted horizon with artificial and real-world forecasts are given by Figures \ref{fig:case5:results_artificial_newLoads}, that we describe in detail, and by Figure \ref{fig:case5:results_volatile_newLoads}, that works analogously.
Generation, storage and line images contain several colored curves that depict the different scenarios; without storage (red), with storage (blue), and storage with variance constraints on generators (green).
Figure \ref{fig:case5:Loads} shows the loads and ten realizations of the uncertain wind generation. Note how the variance grows over time.

Generation and change in generation is given in Figure \ref{fig:case5:Generation}. Without storage (red), the generator needs to provide the difference in power between demand and wind generation. Hence, it reflects the behaviour of the sum of load and wind generation (in this case they have the same behaviour), and assumes all uncertainty of the forecast. 
In contrast, in the scenarios with storage \caseStorage (blue) and additional variance constraints \caseStorageWithVariance (green), the generation curves are almost constant, and do not assume much variance. Looking closely, the variance constraint almost diminishes variance for times $t=3,\dots,9$.
At the end of the horizon, generation curves go down as they have to respond with final storage constraints. 

Storage is depicted in Figure \ref{fig:case5:Storage}. Since there is a surplus of wind generation up to $t=4$, the storage is filled to its limit. Afterwards, the load surpasses generation and the storage empties. Much of the variance is absorbed by the storage;  even more so in scenario \caseStorageWithVariance due to the variance restriction of the generator.

Line flows of all six transmission lines are shown in Figure \ref{fig:case5:LineFlows}. Most obviously, they mirror the loads and uncertain wind generation. 
Without storage, all lines mirror the sum of load and wind generation. 
Upon including storage, lines $1$ and $5$ still mirror the load as they directly connect a generator with a load (see Figure \ref{fig:case5:results_grids}). The other lines are slightly smoothed as they are influenced by the storage.

Replacing the artificial wind forecast with a GPR prediction on real-world data introduces volatility (see Figure \ref{fig:case5:Loads_volatile}).
This leads to a lot more fluctuation for the generators with no storage (see Figure \ref{fig:case5:Generation_volatile}). Including storage leads again to almost constant generation. In terms of storage and line flow there are no differences; the OPF works alike in both trials (see Figures \ref{fig:case5:Storage_volatile} and \ref{fig:case5:LineFlows_volatile}).

Figure \ref{fig:case5:results_grids} visualizes the grids mean values at point $t=4$ in time, for the artificial load, without and with storage (\caseNoStorage and \caseStorage). At this point in time, storage is fully charged and the effect it has on the grids dynamics becomes clearest. 
Figure \ref{fig:case5_results_grids_NoStorage} does not contain storage, while Figure \ref{fig:case5_results_grids_Storage} shows \caseNoStorage with storage.
The effect of storage is that it drastically reduces generation, despite high load.

\begin{figure}
	\centering
	\vspace{3cm}
	
	\begin{subfigure}[c]{\figwidth}
		\centering
		\includegraphics[width=0.45\textwidth]{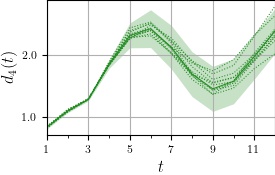}~
		\includegraphics[width=0.45\textwidth]{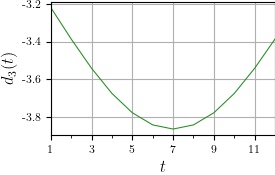}%
		
		\vspace{-2mm}		
		\caption{Left: ten realizations of the uncertain disturbance~$-\rv\load_4(t)$ at bus~$4$ (dots), mean $-\ev{\rv\load_4(t)}$ (solid), and $-(\ev{\rv\load_4(t)} \pm 3\sqrt{\var{\rv\load_4(t)}})$-interval (shaded); right: certain disturbance~$-\load_{3}(t)$ at bus $3$.}
		\label{fig:Disturbances}
	\end{subfigure}
	
	\begin{subfigure}[c]{\figwidth}
		\centering
        \includegraphics[width=0.45\textwidth]{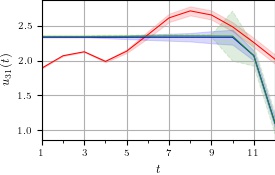}~
        \includegraphics[width=0.45\textwidth]{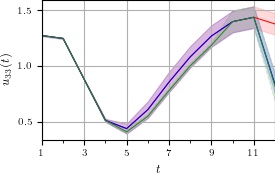}%
		
        \includegraphics[width=0.45\textwidth]{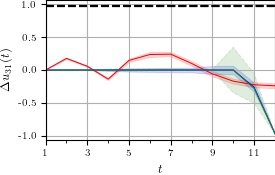}~
        \includegraphics[width=0.45\textwidth]{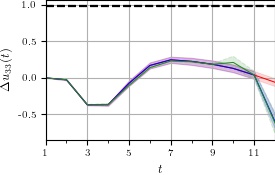}%
        
		\vspace{-2mm}		
		\caption{Upper: Power injections of generators at buses $i \in \{2,4\}$ ; lower: respective change in power injections.}
		\label{fig:Generation}
	\end{subfigure}
	
	\begin{subfigure}[c]{\figwidth}
		\centering
        \includegraphics[width=0.45\textwidth]{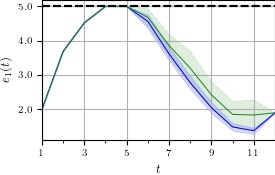}~
        \includegraphics[width=0.45\textwidth]{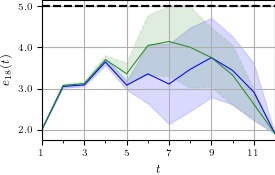}%

		\vspace{-2mm}
		\caption{Upper: Power injections of storages at buses $i \in \{1,4\}$; lower: respective state of storage.}
		\label{fig:Storage}
	\end{subfigure}
	
	\begin{subfigure}[c]{\figwidth}
	\centering
        \includegraphics[width=0.45\textwidth]{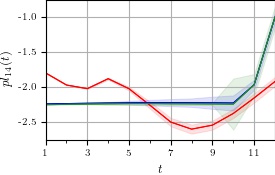}~
        \includegraphics[width=0.45\textwidth]{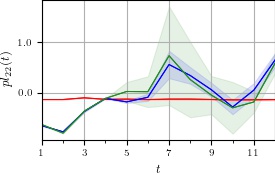}%
        
		\vspace{-2mm}		
		\caption{Line flows across lines $l \in \{14,22\}$.}
		\label{fig:LineFlows}
	\end{subfigure}
	
	\vspace{\adjustlength}
	\caption{\textsc{ieee} 39-bus grid: Results for cases \caseNoStorage (red), \caseStorage (blue), and \caseStorageWithVariance (green). All shown random variables~$\rv{x}$ are depicted in terms of their mean~$\ev{\rv{x}}$ (solid) and the interval $\ev{\rv{x}} \pm \lambda (0.05) \sqrt{\var{\rv{x}}}$ (shaded).}
	\label{fig:IEEEresults}
	\vspace{3cm}
\end{figure}

\begin{figure}
    \centering
    \includegraphics[width=0.5\textwidth]{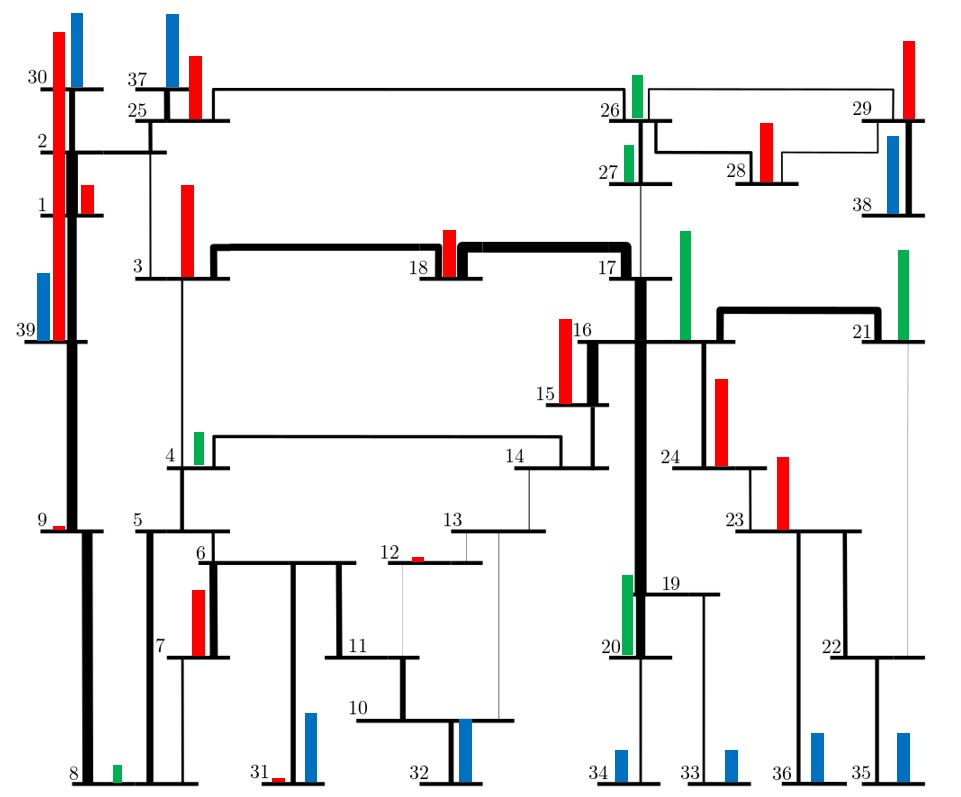}
    \caption{\textsc{ieee} case39: Network state without (upper) and with (lower) storage at time $t=9$, with generation (dark blue), wind generation (light blue), loads (red), storage (green) and line flows (black).}
    \label{fig:case39_grid}
\end{figure}

\subsection{\textsc{ieee} 39-bus test case}
\label{sec:case39}

After having tested method \eqref{eq:SOCP} on a small grid, we show that it works equally well on a larger grid.
The \textsc{ieee} 39-bus system has a total of 10 generators and 46 lines~\cite{Zimmerman11}, see Figure \ref{fig:case39_grid}.
We introduce seven uncertain disturbances at buses~$\mathcal{\Load} = \{4,8,16,20,21,26,27\}$, and five storages are placed at buses~$\mathcal{\Storage} =  \{1,12,14,18,28\}$.
Table~\ref{tab:Parameters} in the Appendix collects all problem-relevant parameters.

In order to check the method and see that storages have the same effect as before, we look at the optimized horizon~$\mathcal{\horizon} = \{1,\dots,12\}$ in Figure \ref{fig:IEEEresults}. The plots are fairly representative for the grid, i.e. the other components behave alike.
Load and wind generation, Figure \ref{fig:Disturbances}, only differ in size, as they are adjusted to the network parameters.
Generation, storage and line flow curves behave similarly.
More components are given in \ref{app:plots}: other loads are equivalent; remaining generators, storages and line flows behave similarly.
Hence, the method also works on this larger grid.

Figure \ref{fig:case39_grid} depicts the grid with all components and line flows. We can see that at time $t=9$ storages are filled and lines adjacent to storage are loaded heavily. Generation is less than in scenario \caseNoStorage without storage.



\subsection{Computational complexity}
\label{sec:complexityAnalysis}

To evaluate the method in terms of scalability, we add \textsc{ieee} cases case57, case118 and case300 to the previous two and perform a complexity analysis with regard to computation time and costs.
Uncertainties are placed at the nodes with the highest load, i.e. the highest impact, and storage systems are placed randomly as placement does not influence computation time. 
We analyse the role of the network size, of the number of uncertain disturbances, of local vs. global balancing, and of storage on computation time. 
Additionally, we show how the costs differ with respect to risk levels, global vs. local balancing as defined in Section \ref{sec:localglobal} and storage. 
\begin{figure}
    \centering
    \includegraphics[width=0.50\textwidth]{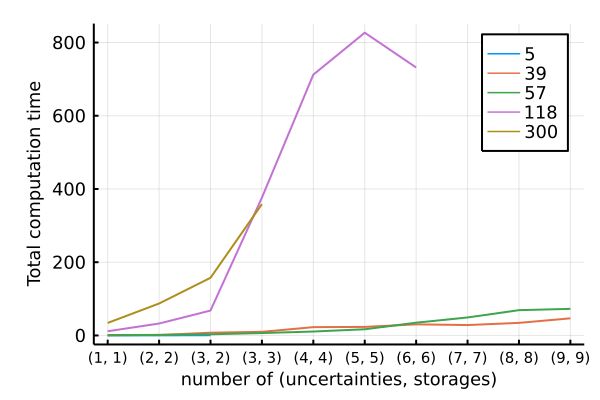}
    \caption{Computation time of all test cases with respect to the number of uncertainties and storages.}
    \label{fig:complexity_all}
\end{figure}

Figure \ref{fig:complexity_all} shows the computational complexity for all cases with one to ten uncertain loads and storage installations. While smaller cases run within seconds, the run time for larger network sizes above 57 rapidly increases to several minutes. We can compute up to 118 nodes efficiently; for a larger number of nodes Mosek runs out of space. Hence, the number of nodes drives computation time up considerably.

\begin{figure}
    \centering
    \includegraphics[width=0.55\textwidth]{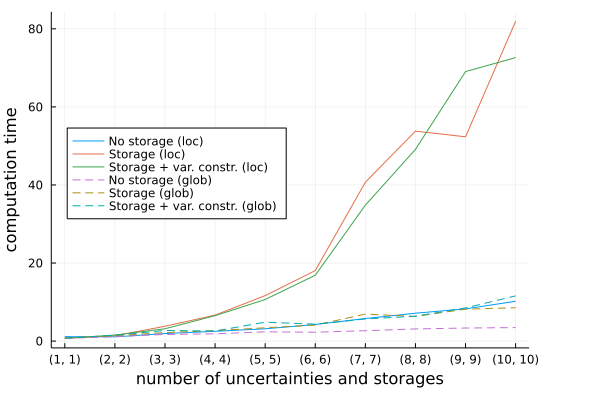}
    \caption{Computation time for case57 of each scenario for local and global balancing with respect to the number of uncertainties and storage.}
    \label{fig:57_complexity_szenarios}
\end{figure}

\begin{figure}
    \includegraphics[width=0.55\textwidth]{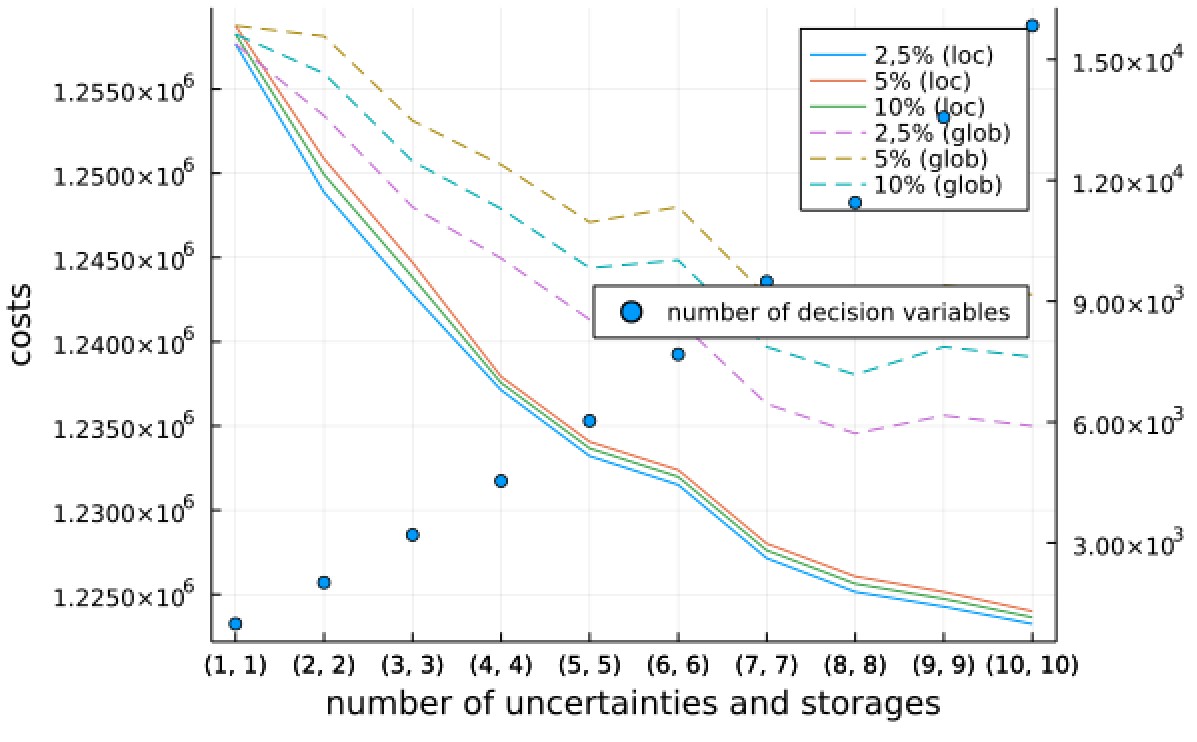}
    \caption{\textsc{ieee} case39: Costs with respect to the different scenarios \caseNoStorage, \caseStorage and \caseStorageWithVariance, different risk levels $\epsilon\in\{2.5\%, 5\%, 10\%\}$ and local vs. global balancing.}
    \label{fig:39_complexity_risk_level}
\end{figure}

We compare the role of different scenarios and local vs. global balancing with the example of case39, in Figure \ref{fig:57_complexity_szenarios}. Clearly, local balancing takes a lot longer than global balancing. Also, storage increases computation time significantly, while adding variance constraints does not, as expected. The number of decision variables (blue points) scales linearly with the number of uncertainties plus storages, as can be seen from equation \eqref{eq:DecisionVariables_local}.
Other cases behave similarly.

Cost is the most interesting measure besides computation time. 
Figure \ref{fig:39_complexity_risk_level} shows the costs for \textsc{ieee} case57 with respect to different risk levels and local vs. global balancing. We can see that the with a growing number of uncertainties and storages the cost decreases. Global balancing seems to be slightly more expensive than local balancing, although looking at the scale values are all close. The different risk levels do not differ much in costs.

\section{Discussion}
\label{sec:discussion}

The main result from Sections \ref{sec:case5} and \ref{sec:case39} is that the method works equally well on various network sizes.
Moreover, we show three outcomes:
(i) Generation profiles are flattened out, hence, generation is a lot more stable with storage in use.
(ii) Costs reduce when more storage and uncertainties are in use, and generation and storage profiles are more similar. This suggests that larger networks can balance out uncertainties better, hence, they are more stable and secure.
(iii) Most of the uncertainty in the wind forecast is absorbed by storages, which means that renewable energy can be well integrated into generation planning, even if there is a lot of uncertainty.

Adding a remark about convergence, we can tell that the network does not converge in several cases: Firstly, when demand is larger than generation, as expected. Secondly, also as expected, when demand is too high in the beginning, because generators cannot ramp up fast enough as they reach their ramp limits.

From Section \ref{sec:complexityAnalysis} testing computation time and costs we can derive five results:
(i) The method is scalable up to roughly 100 nodes without any speedup (e.g. sparsity methods, contraction algorithms).
(ii) Risk levels do not influence costs or computation time.
(iii) local balancing takes a lot longer than global balancing, nevertheless reduces the costs slightly.
(iv) Computation time with respect to the number of uncertainties does not scale linearly with the number of decision variables. 
(v) Storages reduce generation costs notably.
Hence, the method works well on mid-size power grids and is fairly robust with respect to parameter variations. 

Concluding, we can say that the method is robust and performs well on mid-size networks, however, matrix sparcity and contraction algorithms offer large potential for speed-up. Additionally, storage plays a large role in cost reduction, reducing uncertainty by renewables, and stabilizing generation.


\section{Conclusions and Outlook}
\label{sec:conclusion}

We reformulate an intractable optimal power flow problem with uncertain disturbances and chance constraints into a tractable second order cone problem with exact analytical expressions. We modell all decision variables as Gaussian processes and predicted the disturbances with Gaussian process regression.
We test the approach on networks of differing sizes.
The new problem formulation with \gps capturing uncertainty gives realistic results and is computationally efficient for mid-size networks.
The model shows that uncertainty can be handled well by including storage into transmission networks. Almost all uncertainty is absorbed and little left at the generators, which allows for stable generation scheduling. 
Without storage much uncertainty is left at the generators and network control becomes a much more difficult and uncertain task.
Including storage also reduces the cost notably, even with variance constraints.

Further research should aim to adapt the method for practical use. 
As real-world networks are often very large, speeding up the algorithm is a next goal, for example by using the sparsity of matrices.
Also, one can look at non-Gaussian disturbances, or give more detail to the modelling of generators and storage. 
An interesting part will be to automate the Gaussian process regression (GPR) with large amounts of data.


\section{Authors contribution and acknowledgements}

\textbf{Rebecca Bauer (shared first author):} Data Curation, Software: GPR and parts of SOCP, Analysis, Writing: Original Draft, Review \& Editing, Visualization; 
\textbf{Tillmann Mühlpfordt (shared first author):} Conceptualization, Methodology, Software: SOCP, Validation, Analysis, Writing: Original Draft, Visualization, Supervision, Project administration;
\textbf{Nicole Ludwig:} Supervision, Conceptualization, Writing: Original Draft, Review \& Editing;
\textbf{Veit Hagenmeyer:} Supervision, Conceptualization, Review \& Editing, Project administration, Funding acquisition


Rebecca Bauer acknowledges funding by the BMBF-project MOReNet with grant number 05M18CKA.

Nicole Ludwig acknowledges funding by the Deutsche Forschungsgemeinschaft (DFG, German Research Foundation) under Germany’s Excellence Strategy – EXC number 2064/1 – Project number 390727645 and the Athene Grant of the University of Tübingen.


\bibliography{article}


\clearpage

\appendix

\section{Parameter values for case studies}
\label{app:parameters}

\begin{table}[h]
	\centering
	\resizebox{0.85\linewidth}{!}{%
		\centering
		\begin{minipage}{\linewidth}
			\centering
			\caption{Parameter values for case studies.}	
			\label{tab:Parameters}
			\centering	
			\begin{tabular}{p{0.45cm}llll}	
				\toprule
				\multirow{2}{2.75em}{$i \in \mathcal{\Gen}$}   &   $\underline{\gen}_i=0.0$ &   $\overline{\gen}_i =\,1.1 \overline{p}_i$ &       $\Delta \underline{\gen}_i =-0.15\overline{p}_i$ &        $\Delta \overline{\gen}_i =0.15\overline{p}_i$\\
				&      $\gamma_{i,2} = 0.01$ &      $\gamma_{i,1}=0.3$ &              $\gamma_{i,0} =0.2$ &  \\
				\midrule
				\multirow{2}{2.75em}{$i \in \mathcal{\Storage}$} &  $\underline{\energy}_i =0.0$ &  $
				\overline{\energy}_i =6.0$ &         $\underline{\storage}_i  =-10.0$ &          $\overline{\storage}_i = 10.0$\\
				& $\underline{\energy}_i^\horizon = 0.19 $ & $\overline{\energy}_i^\horizon =0.21 $ & $\ev{\rv\energy_i^{\text{\textsc{ic}}}} =2.0 $ & $\var{\rv\energy_i^{\text{\textsc{ic}}}} = 0.0$\\
				\midrule
				\multirow{1}{2.75em}{$j \in \mathcal{L}$}     & $\underline{\lineflow}_j=-0.85\overline{p}_{l,j}$ & $\overline{\lineflow}_j = 0.85 \overline{p}_{l,j}$ &  \multicolumn{2}{|c}{$\overline{p}_i$, $\overline{p}_{l,j}$ taken from case file \cite{Zimmerman11}}\\ \bottomrule
			\end{tabular}
		\end{minipage}
	}
	\vspace{\adjustlength}
\end{table}

\vspace{5mm}

\clearpage

\section{Additional plots of case studies}
\label{app:plots}






\begin{figure}[h]
	\centering
    
    \begin{subfigure}[c]{\figwidth}
    \centering
        \includegraphics[width=0.45\textwidth]{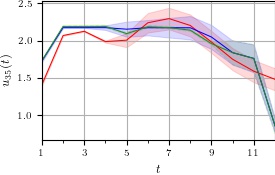}~
    	\includegraphics[width=0.45\textwidth]{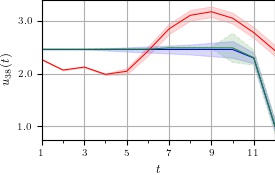}%
    	\vspace{-2mm}	
    	\caption{Power injections of generators at buses $i \in \{6,9\}$.}
    \end{subfigure}
    
    \begin{subfigure}[c]{\figwidth}
    \centering
    	\includegraphics[width=0.45\textwidth]{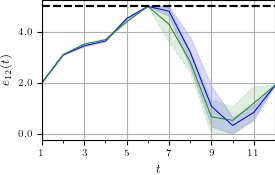}~
    	\includegraphics[width=0.45\textwidth]{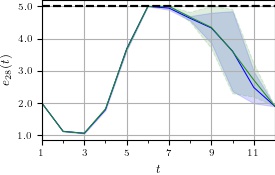}%
    	\vspace{-2mm}	
    	\caption{Power injections of storages at buses $i \in \{2,5\}$.}
    \end{subfigure}
    
    \begin{subfigure}[c]{\figwidth}
    \centering
        \includegraphics[width=0.45\textwidth]{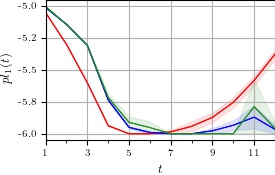}~
    	\includegraphics[width=0.45\textwidth]{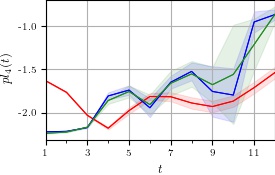}%
    \end{subfigure}
    
    \begin{subfigure}[c]{\figwidth}
    \centering
    	\includegraphics[width=0.45\textwidth]{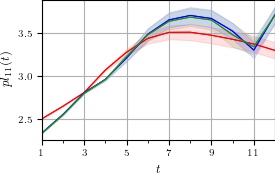}~
    	\includegraphics[width=0.45\textwidth]{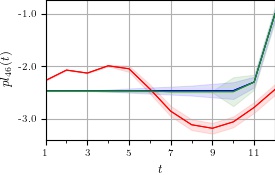}%
    	\vspace{-2mm}	
    	\caption{Power flows at lines $i\in\{1,4,11,46\}$.}
    \end{subfigure}
    	
    \vspace{2mm}	
	\vspace{\adjustlength}
	\caption{\textsc{ieee} 39-bus grid: Results for 7 uncertainties and 5 storage systems for cases \caseNoStorage (red), \caseStorage (blue), and \caseStorageWithVariance (green). All shown random variables~$\rv{x}$ are depicted in terms of their mean~$\ev{\rv{x}}$ (solid) and the interval $\ev{\rv{x}} \pm \lambda (0.05) \sqrt{\var{\rv{x}}}$ (shaded).}
\end{figure}


\begin{figure}
	\centering
    
    \begin{subfigure}[c]{\figwidth}
    \centering
        \includegraphics[width=0.45\textwidth]{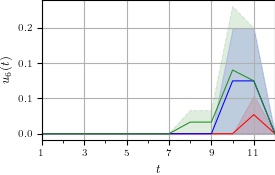}~
    	\includegraphics[width=0.45\textwidth]{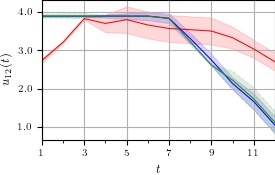}%
    	\vspace{-2mm}	
    	\caption{Power injections of generators at buses $i \in \{4,7\}$.}
    \end{subfigure}
    
    \begin{subfigure}[c]{\figwidth}
    \centering
    	\includegraphics[width=0.45\textwidth]{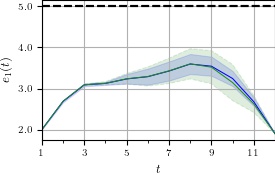}~
    	\includegraphics[width=0.45\textwidth]{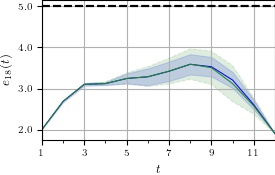}%
    	\vspace{-2mm}	
    	\caption{Power injections of storages at buses $i \in \{1,4\}$.}
    \end{subfigure}
    
    \begin{subfigure}[c]{\figwidth}
    \centering
        \includegraphics[width=0.45\textwidth]{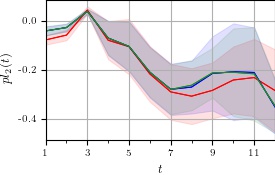}~
    	\includegraphics[width=0.45\textwidth]{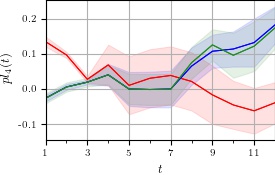}%
    \end{subfigure}
    \begin{subfigure}[c]{\figwidth}
    \centering
    	\includegraphics[width=0.45\textwidth]{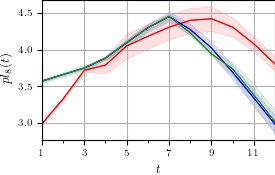}~
    	\includegraphics[width=0.45\textwidth]{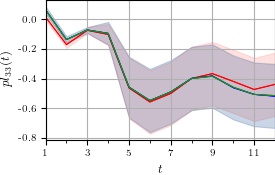}%
    	\vspace{-2mm}	
    	\caption{Power flows at lines $i\in\{2,4,8,33\}$.}
    \end{subfigure}
    	
    \vspace{2mm}	
	\vspace{\adjustlength}
	\caption{\textsc{ieee} 57-bus grid: Results for 7 uncertainties and 5 storage systems for cases \caseNoStorage (red), \caseStorage (blue), and \caseStorageWithVariance (green). All shown random variables~$\rv{x}$ are depicted in terms of their mean~$\ev{\rv{x}}$ (solid) and the interval $\ev{\rv{x}} \pm \lambda (0.05) \sqrt{\var{\rv{x}}}$ (shaded).}
\end{figure}

\newpage

\begin{figure}
	\centering
	
    \begin{subfigure}[c]{\figwidth}
    \centering
        \includegraphics[width=0.45\textwidth]{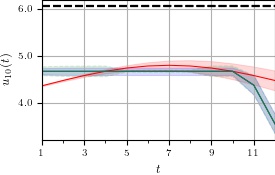}~
    	\includegraphics[width=0.45\textwidth]{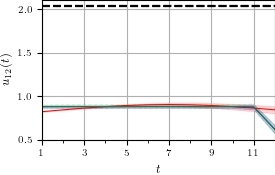}%
    	
    	\includegraphics[width=0.45\textwidth]{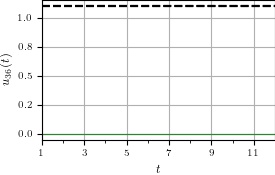}~
    	\includegraphics[width=0.45\textwidth]{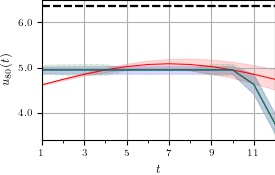}%
    	\vspace{-2mm}	
    	\caption{Power injections of generators at buses $i \in \{5,6,17,37\}$.}
    \end{subfigure}
    
    
    \begin{subfigure}[c]{\figwidth}
    \centering
    	\includegraphics[width=0.45\textwidth]{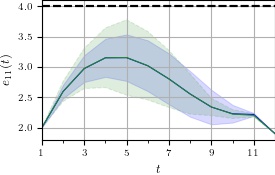}~
    	\includegraphics[width=0.45\textwidth]{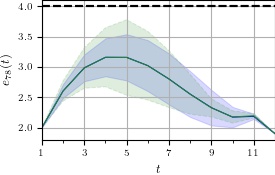}%
    	\vspace{-2mm}	
    	\caption{Power injections of storages at buses $i \in \{1,4\}$.}
    \end{subfigure}
    
    
    
    \vspace{2mm}	
	\vspace{\adjustlength}
	\caption{\textsc{ieee} 118-bus grid: Results for 10 uncertainties and 5 storage systems cases \caseNoStorage (red), \caseStorage (blue), and \caseStorageWithVariance (green). All shown random variables~$\rv{x}$ are depicted in terms of their mean~$\ev{\rv{x}}$ (solid) and the interval $\ev{\rv{x}} \pm \lambda (0.05) \sqrt{\var{\rv{x}}}$ (shaded).}
\end{figure}

\begin{figure}
	\centering
	
    \begin{subfigure}[c]{\figwidth}
        \centering
        \includegraphics[width=0.45\textwidth]{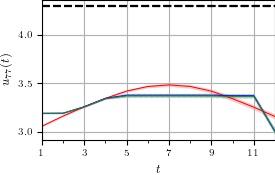}~
    	\includegraphics[width=0.45\textwidth]{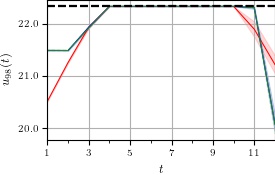}%
    	
        \includegraphics[width=0.45\textwidth]{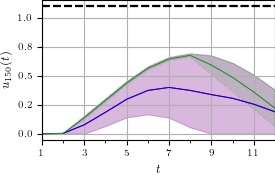}~
    	\includegraphics[width=0.45\textwidth]{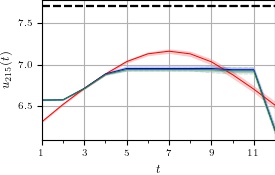}%
    	\vspace{-2mm}	
    	\caption{Power injections of generators at buses $i \in \{8,11,24,40\}$.}
    \end{subfigure}
    
    
    \begin{subfigure}[c]{\figwidth}
    \centering
    	\includegraphics[width=0.45\textwidth]{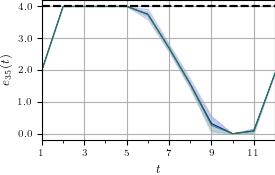}~
    	\includegraphics[width=0.45\textwidth]{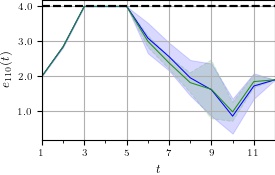}%
    	\vspace{-2mm}	
    	\caption{Power injections of storages at buses $i \in \{1,4\}$}
    \end{subfigure}
    
    
    \vspace{2mm}	
	\vspace{\adjustlength}
	\caption{\textsc{ieee} 300-bus grid: Results for 4 uncertainties and 4 storage systems for cases \caseNoStorage (red), \caseStorage (blue), and \caseStorageWithVariance (green). All shown random variables~$\rv{x}$ are depicted in terms of their mean~$\ev{\rv{x}}$ (solid) and the interval $\ev{\rv{x}} \pm \lambda (0.05) \sqrt{\var{\rv{x}}}$ (shaded).}
\end{figure}

\end{document}

%% file: macros.tex
\usepackage{amssymb}
\usepackage{amsmath}
\usepackage{amsthm}
\usepackage[english]{babel}
\usepackage{xspace}
\usepackage{cases}
\usepackage{color}
\usepackage{graphicx}
\usepackage[font=footnotesize]{caption}
\usepackage[font=footnotesize]{subcaption}
\usepackage{epstopdf}
\usepackage{upgreek}
\usepackage{booktabs}
\usepackage{array}
\usepackage{blindtext}
\usepackage{multirow}
\usepackage{bbold}
\usepackage[hidelinks]{hyperref}
\usepackage{soul}
\usepackage{enumerate}
\usepackage{blindtext}
\usepackage{graphicx}
\usepackage{xcolor}
\usepackage{siunitx}
\usepackage{pgf}
\usepackage{lipsum}
\usepackage[normalem]{ulem}
\usepackage{glossaries}

\newcommand{\figwidth}{\linewidth}

\newcommand{\nbus}{N}

\newcommand{\nline}{N_l}

\newcommand{\horizon}{T}
\newcommand{\pnet}{p}
\newcommand{\gen}{u}
\newcommand{\load}{d}

\newcommand{\storage}{s}
\newcommand{\energy}{e}
\newcommand{\lineflow}{c}

\newcommand{\caseNoStorage}{\textsc{S}1\xspace}
\newcommand{\caseStorage}{\textsc{S}2\xspace}
\newcommand{\caseStorageWithVariance}{\textsc{S}3\xspace}

\newcommand{\Load}{\MakeUppercase{\load}}
\newcommand{\Gen}{\MakeUppercase{\gen}}
\newcommand{\Storage}{\MakeUppercase{\storage}}

\newcommand{\ptdfmat}{\Phi}

\newcommand{\adjustlength}{-4.5mm}

\newacronym{gp}{GP}{Gaussian process}

\newcommand{\gps}{\textsc{gp}s\xspace}
\newcommand{\opf}{\textsc{opf}\xspace}

\newcommand{\dc}{\textsc{dc}\xspace}

\newcommand{\socp}{\textsc{socp}\xspace}

\newcommand{\rv}[1]{\mathsf{#1}}

\newcommand{\ev}[1]{\mathbb{E}(#1)}

\newcommand{\var}[1]{\mathbb{V}(#1)}

\newcommand{\prob}[1]{\mathbb{P}( #1 )}

\newcommand{\pvar}{u}
\newcommand{\pvarup}{\expandafter\uppercase\expandafter{\pvar}}

\newcommand{\pfix}{d}
\newcommand{\pfixup}{\expandafter\uppercase\expandafter{\pfix}}

\definecolor{till}{rgb}{.1,.4,.9}
\definecolor{timm}{rgb}{1.0,.0,1.0} 
\definecolor{veit}{rgb}{0.,.8,.4}
\definecolor{revised}{rgb}{.2,.8,.1}

\newacronym{ess}{ESS}{energy storage system}
\newacronym{cc}{CC}{chance-constrained}
\newacronym{opf}{OPF}{optimal power flow}
\newacronym{socp}{SOCP}{second-order cone problem}
\newacronym{res}{RES}{renewable energy source}

%% file: article.bbl
\begin{thebibliography}{10}
\expandafter\ifx\csname url\endcsname\relax
  \def\url#1{\texttt{#1}}\fi
\expandafter\ifx\csname urlprefix\endcsname\relax\def\urlprefix{URL }\fi
\expandafter\ifx\csname href\endcsname\relax
  \def\href#1#2{#2} \def\path#1{#1}\fi

\bibitem{Capitanescu12}
F.~Capitanescu, S.~Fliscounakis, P.~Panciatici, L.~Wehenkel, Cautious operation
  planning under uncertainties, {IEEE} Transactions on Power Systems 27~(4)
  (2012) 1859--1869.
\newblock \href {http://dx.doi.org/10.1109/tpwrs.2012.2188309}
  {\path{doi:10.1109/tpwrs.2012.2188309}}.

\bibitem{Vrakopoulou17}
M.~Vrakopoulou, I.~Hiskens, Optimal policy-based control of generation and
  {HVDC} lines in power systems under uncertainty, {IEEE} Manchester
  PowerTech(2017) \href {http://dx.doi.org/10.1109/ptc.2017.7981262}
  {\path{doi:10.1109/ptc.2017.7981262}}.

\bibitem{Roald18}
L.~Roald, G.~Andersson, Chance-constrained {AC} optimal power flow:
  Reformulations and efficient algorithms, {IEEE} Transactions on Power Systems
  33~(3) (2018) 2906--2918.
\newblock \href {http://dx.doi.org/10.1109/TPWRS.2017.2745410}
  {\path{doi:10.1109/TPWRS.2017.2745410}}.

\bibitem{Guo18}
Y.~Guo, K.~Baker, E.~Dall{\textquotesingle}Anese, Z.~Hu, T.~Summers, Data-based
  distributionally robust stochastic optimal power flow, part i: Methodologies,
  {IEEE} Transactions on Power Systems (2018) 1--1\href
  {http://dx.doi.org/10.1109/tpwrs.2018.2878385}
  {\path{doi:10.1109/tpwrs.2018.2878385}}.

\bibitem{Li17}
B.~Li, M.~Vrakopoulou, J.~Mathieu, Chance constrained reserve scheduling using
  uncertain controllable loads part {II}: Analytical reformulation, {IEEE}
  Transactions on Smart Grid (2017) 1--1\href
  {http://dx.doi.org/10.1109/tsg.2017.2773603}
  {\path{doi:10.1109/tsg.2017.2773603}}.

\bibitem{Warrington13}
J.~Warrington, P.~Goulart, S.~Mari\'ethoz, M.~Morari, Policy-based reserves for
  power systems, IEEE Transactions on Power Systems 28~(4) (2013) 4427--4437.
\newblock \href {http://dx.doi.org/10.1109/TPWRS.2013.2269804}
  {\path{doi:10.1109/TPWRS.2013.2269804}}.

\bibitem{molina_distributed_2012}
M.~G. Molina, Distributed energy storage systems for applications in future
  smart grids (2012) 1--7\href {http://dx.doi.org/10.1109/TDC-LA.2012.6319051}
  {\path{doi:10.1109/TDC-LA.2012.6319051}}.

\bibitem{figgener_development_2021}
J.~Figgener, P.~Stenzel, K.-P. Kairies, J.~Linßen, D.~Haberschusz, O.~Wessels,
  M.~Robinius, D.~Stolten, D.~U. Sauer, The development of stationary battery
  storage systems in germany – status 2020 33 (2021) 101982.
\newblock \href {http://dx.doi.org/10.1016/j.est.2020.101982}
  {\path{doi:10.1016/j.est.2020.101982}}.

\bibitem{zsiboracs_intermittent_2019}
H.~Zsiborács, N.~H. Baranyai, A.~Vincze, L.~Zentkó, Z.~Birkner, K.~Máté,
  G.~Pintér, Intermittent renewable energy sources: The role of energy storage
  in the european power system of 2040 8~(7) (2019) 729.
\newblock \href {http://dx.doi.org/10.3390/electronics8070729}
  {\path{doi:10.3390/electronics8070729}}.

\bibitem{gulagi_role_2018}
A.~Gulagi, D.~Bogdanov, C.~Breyer, The role of storage technologies in energy
  transition pathways towards achieving a fully sustainable energy system for
  india 17 (2018) 525--539.
\newblock \href {http://dx.doi.org/10.1016/j.est.2017.11.012}
  {\path{doi:10.1016/j.est.2017.11.012}}.

\bibitem{keck_impact_2019}
F.~Keck, M.~Lenzen, A.~Vassallo, M.~Li, The impact of battery energy storage
  for renewable energy power grids in australia 173 (2019) 647--657.
\newblock \href {http://dx.doi.org/10.1016/j.energy.2019.02.053}
  {\path{doi:10.1016/j.energy.2019.02.053}}.

\bibitem{wood_power_2014}
A.~J. Wood, B.~F. Wollenberg, G.~B. Sheblé, Power Generation, Operation and
  Control, Wiley, 2014.

\bibitem{Muehlpfordt18c}
T.~M{\"u}hlpfordt, T.~Faulwasser, V.~Hagenmeyer, A generalized framework for
  chance-constrained optimal power flow, Sustainable Energy, Grids and Networks
  16 (2018) 231 -- 242.
\newblock \href {http://dx.doi.org/https://doi.org/10.1016/j.segan.2018.08.002}
  {\path{doi:https://doi.org/10.1016/j.segan.2018.08.002}}.

\bibitem{Hong16}
T.~Hong, P.~Pinson, S.~Fan, H.~Zareipour, A.~Troccoli, R.~Hyndman,
  Probabilistic energy forecasting: Global energy forecasting competition 2014
  and beyond, International Journal of Forecasting 32~(3) (2016) 896 -- 913.
\newblock \href
  {http://dx.doi.org/https://doi.org/10.1016/j.ijforecast.2016.02.001}
  {\path{doi:https://doi.org/10.1016/j.ijforecast.2016.02.001}}.

\bibitem{quatile_zhang_2020}
Y.~Zhang, Y.~Zhao, G.~Pan, J.~Zhang, Wind speed interval prediction based on
  lorenz disturbance distribution, IEEE Transactions on Sustainable Energy
  11~(2) (2020) 807--816.
\newblock \href {http://dx.doi.org/10.1109/TSTE.2019.2907699}
  {\path{doi:10.1109/TSTE.2019.2907699}}.

\bibitem{quantile_solar_2017}
P.~Lauret, M.~David, H.~T.~C. Pedro, Probabilistic solar forecasting using
  quantile regression models, Energies 10~(10)(2017) .
\newblock \href {http://dx.doi.org/10.3390/en10101591}
  {\path{doi:10.3390/en10101591}}.

\bibitem{Ding16}
H.~Ding, P.~Pinson, Z.~Hu, Y.~Song, Optimal offering and operating strategies
  for wind-storage systems with linear decision rules, {IEEE} Transactions on
  Power Systems 31~(6) (2016) 4755--4764.
\newblock \href {http://dx.doi.org/10.1109/tpwrs.2016.2521177}
  {\path{doi:10.1109/tpwrs.2016.2521177}}.

\bibitem{Bucher17}
M.~Bucher, M.~Ortega-Vazquez, D.~Kirschen, G.~Andersson, Robust allocation of
  reserves considering different reserve types and the flexibility from {HVDC},
  {IET} Generation, Transmission {\&} Distribution 11~(6) (2017) 1472--1478.
\newblock \href {http://dx.doi.org/10.1049/iet-gtd.2016.1014}
  {\path{doi:10.1049/iet-gtd.2016.1014}}.

\bibitem{Fabietti17}
L.~Fabietti, T.~Gorecki, E.~Namor, F.~Sossan, M.~Paolone, C.~Jones, Dispatching
  active distribution networks through electrochemical storage systems and
  demand side management (2017) 1241--1247\href
  {http://dx.doi.org/10.1109/CCTA.2017.8062629}
  {\path{doi:10.1109/CCTA.2017.8062629}}.

\bibitem{Fabietti18}
L.~Fabietti, T.~Gorecki, F.~Qureshi, A.~Bitlislio\u{g}lu, I.~Lymperopoulos,
  C.~Jones, Experimental implementation of frequency regulation services using
  commercial buildings, {IEEE} Transactions on Smart Grid 9~(3) (2018)
  1657--1666.
\newblock \href {http://dx.doi.org/10.1109/TSG.2016.2597002}
  {\path{doi:10.1109/TSG.2016.2597002}}.

\bibitem{Vrakopoulou13b}
M.~Vrakopoulou, K.~Margellos, J.~Lygeros, G.~Andersson, A probabilistic
  framework for reserve scheduling and n-1 security assessment of systems with
  high wind power penetration, IEEE Transactions on Power Systems 28~(4) (2013)
  3885--3896.
\newblock \href {http://dx.doi.org/10.1109/TPWRS.2013.2272546}
  {\path{doi:10.1109/TPWRS.2013.2272546}}.

\bibitem{Louca16}
R.~Louca, E.~Bitar, Stochastic {AC} optimal power flow with affine
  recourse(2016) \href {http://dx.doi.org/10.1109/cdc.2016.7798626}
  {\path{doi:10.1109/cdc.2016.7798626}}.

\bibitem{MunozAlvarez14}
D.~Munoz-Alvarez, E.~Bitar, L.~Tong, J.~Wang, Piecewise affine dispatch
  policies for economic dispatch under uncertainty(2014) \href
  {http://dx.doi.org/10.1109/pesgm.2014.6939369}
  {\path{doi:10.1109/pesgm.2014.6939369}}.

\bibitem{Oldewurtel12}
F.~Oldewurtel, A.~Parisio, C.~Jones, D.~Gyalistras, M.~Gwerder, V.~Stauch,
  B.~Lehmann, M.~Morari, Use of model predictive control and weather forecasts
  for energy efficient building climate control, Energy and Buildings 45 (2012)
  15--27.
\newblock \href {http://dx.doi.org/10.1016/j.enbuild.2011.09.022}
  {\path{doi:10.1016/j.enbuild.2011.09.022}}.

\bibitem{Muehlpfordt17a}
T.~M{\"u}hlpfordt, T.~Faulwasser, L.~Roald, V.~Hagenmeyer, Solving optimal
  power flow with non-gaussian uncertainties via polynomial chaos expansion
  (2017) 4490--4496\href {http://dx.doi.org/10.1109/CDC.2017.8264321}
  {\path{doi:10.1109/CDC.2017.8264321}}.

\bibitem{Muehlpfordt16b}
T.~M{\"u}hlpfordt, T.~Faulwasser, V.~Hagenmeyer, Solving stochastic {AC} power
  flow via polynomial chaos expansion (2016) 70--76\href
  {http://dx.doi.org/10.1109/CCA.2016.7587824}
  {\path{doi:10.1109/CCA.2016.7587824}}.

\bibitem{Vrakopoulou13}
M.~Vrakopoulou, M.~Katsampani, K.~Margellos, J.~Lygeros, G.~Andersson,
  Probabilistic security-constrained {AC} optimal power flow (2013) 12--6\href
  {http://dx.doi.org/10.1109/PTC.2013.6652374}
  {\path{doi:10.1109/PTC.2013.6652374}}.

\bibitem{Vrakopoulou17b}
M.~Vrakopoulou, B.~Li, J.~Mathieu, Chance constrained reserve scheduling using
  uncertain controllable loads part i: Formulation and scenario-based analysis,
  {IEEE} Transactions on Smart Grid (2017) 1--1\href
  {http://dx.doi.org/10.1109/tsg.2017.2773627}
  {\path{doi:10.1109/tsg.2017.2773627}}.

\bibitem{hemmati_stochastic_2016}
R.~Hemmati, H.~Saboori, S.~Saboori, Stochastic risk-averse coordinated
  scheduling of grid integrated energy storage units in transmission
  constrained wind-thermal systems within a conditional value-at-risk framework
  113 (2016) 762--775.
\newblock \href {http://dx.doi.org/10.1016/j.energy.2016.07.089}
  {\path{doi:10.1016/j.energy.2016.07.089}}.

\bibitem{ayyagari_chance_2017}
K.~S. Ayyagari, N.~Gatsis, A.~F. Taha, Chance constrained optimization of
  distributed energy resources via affine policies (2017) 1050--1054\href
  {http://dx.doi.org/10.1109/GlobalSIP.2017.8309121}
  {\path{doi:10.1109/GlobalSIP.2017.8309121}}.

\bibitem{summers_stochastic_2015}
T.~Summers, J.~Warrington, M.~Morari, J.~Lygeros, Stochastic optimal power flow
  based on conditional value at risk and distributional robustness {\textbar}
  elsevier enhanced reader 72 (2015) 116--125.
\newblock \href {http://dx.doi.org/10.1016/j.ijepes.2015.02.024}
  {\path{doi:10.1016/j.ijepes.2015.02.024}}.

\bibitem{li_analytical_2015}
B.~Li, J.~L. Mathieu, Analytical reformulation of chance-constrained optimal
  power flow with uncertain load control (2015) 1--6\href
  {http://dx.doi.org/10.1109/PTC.2015.7232803}
  {\path{doi:10.1109/PTC.2015.7232803}}.

\bibitem{bienstock_chance_2012}
D.~Bienstock, M.~Chertkov, S.~Harnett, Chance constrained optimal power flow:
  Risk-aware network control under uncertainty 56(2012) .
\newblock \href {http://dx.doi.org/10.1137/130910312}
  {\path{doi:10.1137/130910312}}.

\bibitem{roald_analytical_2013}
L.~Roald, F.~Oldewurtel, T.~Krause, G.~Andersson, Analytical reformulation of
  security constrained optimal power flow with probabilistic constraints (2013)
  1--6\href {http://dx.doi.org/10.1109/PTC.2013.6652224}
  {\path{doi:10.1109/PTC.2013.6652224}}.

\bibitem{sjodin_risk-mitigated_2012}
E.~Sjödin, D.~F. Gayme, U.~Topcu, Risk-mitigated optimal power flow for wind
  powered grids (2012) 4431--4437{ISSN}: 2378-5861.
\newblock \href {http://dx.doi.org/10.1109/ACC.2012.6315377}
  {\path{doi:10.1109/ACC.2012.6315377}}.

\bibitem{zhang_distributionally_2017}
Y.~Zhang, S.~Shen, J.~L. Mathieu, Distributionally robust chance-constrained
  optimal power flow with uncertain renewables and uncertain reserves provided
  by loads 32~(2) (2017) 1378--1388.
\newblock \href {http://dx.doi.org/10.1109/TPWRS.2016.2572104}
  {\path{doi:10.1109/TPWRS.2016.2572104}}.

\bibitem{xie_distributionally_2018}
W.~Xie, S.~Ahmed, Distributionally robust chance constrained optimal power flow
  with renewables: A conic reformulation 33~(2) (2018) 1860--1867.
\newblock \href {http://dx.doi.org/10.1109/TPWRS.2017.2725581}
  {\path{doi:10.1109/TPWRS.2017.2725581}}.

\bibitem{mitrentsis_probabilistic_2021}
G.~Mitrentsis, H.~Lens, Probabilistic dynamic model of active distribution
  networks using gaussian processes (2021) 1--6\href
  {http://dx.doi.org/10.1109/PowerTech46648.2021.9494816}
  {\path{doi:10.1109/PowerTech46648.2021.9494816}}.

\bibitem{Roberts12}
S.~Roberts, M.~Osborne, M.~Ebden, S.~Reece, N.~Gibson, S.~Aigrain, Gaussian
  processes for time-series modelling, Philosophical Transactions of the Royal
  Society A: Mathematical, Physical and Engineering Sciences 371~(1984) (2012)
  20110550--20110550.
\newblock \href {http://dx.doi.org/10.1098/rsta.2011.0550}
  {\path{doi:10.1098/rsta.2011.0550}}.

\bibitem{Kou13}
P.~Kou, F.~Gao, X.~Guan, Sparse online warped gaussian process for wind power
  probabilistic forecasting, Applied Energy 108 (2013) 410--428.
\newblock \href {http://dx.doi.org/10.1016/j.apenergy.2013.03.038}
  {\path{doi:10.1016/j.apenergy.2013.03.038}}.

\bibitem{Chen14}
N.~Chen, Z.~Qian, I.~Nabney, X.~Meng, Wind power forecasts using gaussian
  processes and numerical weather prediction, {IEEE} Transactions on Power
  Systems 29~(2) (2014) 656--665.
\newblock \href {http://dx.doi.org/10.1109/tpwrs.2013.2282366}
  {\path{doi:10.1109/tpwrs.2013.2282366}}.

\bibitem{Sheng18}
H.~Sheng, J.~Xiao, Y.~Cheng, Q.~Ni, S.~Wang, Short-term solar power forecasting
  based on weighted gaussian process regression, {IEEE} Transactions on
  Industrial Electronics 65~(1) (2018) 300--308.
\newblock \href {http://dx.doi.org/10.1109/tie.2017.2714127}
  {\path{doi:10.1109/tie.2017.2714127}}.

\bibitem{Mori08}
H.~Mori, M.~Ohmi, Probabilistic short-term load forecasting with gaussian
  processes (2008) 452--457\href {http://dx.doi.org/10.1109/isap.2005.1599306}
  {\path{doi:10.1109/isap.2005.1599306}}.

\bibitem{Leith04}
D.~Leith, M.~Heidl, J.~Ringwood, Gaussian process prior models for electrical
  load forecasting (2004) 112--117\href
  {http://dx.doi.org/10.1109/PMAPS.2004.242921}
  {\path{doi:10.1109/PMAPS.2004.242921}}.

\bibitem{Lloyd14}
J.~Lloyd, {GEFCom}2012 hierarchical load forecasting: Gradient boosting
  machines and gaussian processes, International Journal of Forecasting 30~(2)
  (2014) 369--374.
\newblock \href {http://dx.doi.org/10.1016/j.ijforecast.2013.07.002}
  {\path{doi:10.1016/j.ijforecast.2013.07.002}}.

\bibitem{Rogers11}
A.~Rogers, S.~Maleki, S.~Ghosh, N.~Jennings, Adaptive home heating control
  through gaussian process prediction and mathematical programming (2011)
  71--78.

\bibitem{Blum13}
M.~Blum, M.~Riedmiller, Electricity demand forecasting using gaussian processes
  (2013) 10--13.

\bibitem{McLoughlin13}
F.~McLoughlin, A.~Duffy, M.~Conlon, Evaluation of time series techniques to
  characterise domestic electricity demand, Energy 50 (2013) 120--130.
\newblock \href {http://dx.doi.org/10.1016/j.energy.2012.11.048}
  {\path{doi:10.1016/j.energy.2012.11.048}}.

\bibitem{Rasmussen06}
C.~Rasmussen, C.~Williams, Gaussian processes for machine learning (2006) 248.

\bibitem{horsch_linear_2017}
J.~Hörsch, H.~Ronellenfitsch, D.~Witthaut, T.~Brown, Linear optimal power flow
  using cycle flows 158(2017) .
\newblock \href {http://dx.doi.org/10.1016/j.epsr.2017.12.034}
  {\path{doi:10.1016/j.epsr.2017.12.034}}.

\bibitem{duvenaud_automatic_nodate}
D.~K. Duvenaud, \href{https://www.cs.toronto.edu/~duvenaud/cookbook/}{Automatic
  model construction with gaussian processes} (2020) 157.
\newline\urlprefix\url{https://www.cs.toronto.edu/~duvenaud/cookbook/}

\bibitem{Bienstock14}
D.~Bienstock, M.~Chertkov, S.~Harnett,
  \href{http://dx.doi.org/10.1137/130910312}{Chance-constrained optimal power
  flow: Risk-aware network control under uncertainty}, {SIAM} Review 56~(3)
  (2014) 461--495.
\newblock \href {http://arxiv.org/abs/http://dx.doi.org/10.1137/130910312}
  {\path{arXiv:http://dx.doi.org/10.1137/130910312}}, \href
  {http://dx.doi.org/10.1137/130910312} {\path{doi:10.1137/130910312}}.
\newline\urlprefix\url{http://dx.doi.org/10.1137/130910312}

\bibitem{de_felice_matteo_2021_4682697}
M.~De~Felice, {ENTSO-E Actual Generation of Wind units: data from 21-12-2014 to
  11-04-2021}(2021) \href {http://dx.doi.org/10.5281/zenodo.4682697}
  {\path{doi:10.5281/zenodo.4682697}}.

\bibitem{10.5555/1593511}
G.~Van~Rossum, F.~L. Drake, Python 3 Reference Manual, CreateSpace, Scotts
  Valley, CA, (2009).

\bibitem{GPflow2017}
A.~G. d.~G. Matthews, M.~{van der Wilk}, T.~Nickson, K.~Fujii,
  A.~{Boukouvalas}, P.~{Le{\'o}n-Villagr{\'a}}, Z.~Ghahramani, J.~Hensman,
  \href{http://jmlr.org/papers/v18/16-537.html}{{{GP}flow: A {G}aussian process
  library using {T}ensor{F}low}}, Journal of Machine Learning Research 18~(40)
  (2017) 1--6.
\newline\urlprefix\url{http://jmlr.org/papers/v18/16-537.html}

\bibitem{Bezanson2017}
J.~Bezanson, A.~Edelman, S.~Karpinski, V.~Shah, Julia: A fresh approach to
  numerical computing, {SIAM} Review 59~(1) (2017) 65--98.
\newblock \href {http://dx.doi.org/10.1137/141000671}
  {\path{doi:10.1137/141000671}}.

\bibitem{Dunning2017}
I.~Dunning, J.~Huchette, M.~Lubin, {JuMP}: A modeling language for mathematical
  optimization, {SIAM} Review 59~(2) (2017) 295--320.
\newblock \href {http://dx.doi.org/10.1137/15m1020575}
  {\path{doi:10.1137/15m1020575}}.

\bibitem{muhlpfordt_git_nodate}
T.~Mühlpfordt, R.~Bauer, \href{https://github.com/KIT-IAI/DCsOPF}{Git
  repository {KIT}-{IAI}/{DCsOPF}}(2022) .
\newline\urlprefix\url{https://github.com/KIT-IAI/DCsOPF}

\bibitem{Zimmerman11}
R.~Zimmerman, C.~Murillo-Sanchez, R.~Thomas, {MATPOWER}: Steady-state
  operations, planning, and analysis tools for power systems research and
  education, {IEEE} Transactions on Power Systems 26~(1) (2011) 12--19.
\newblock \href {http://dx.doi.org/10.1109/TPWRS.2010.2051168}
  {\path{doi:10.1109/TPWRS.2010.2051168}}.

\end{thebibliography}
